\documentclass[journal,draftclsnofoot,onecolumn,12pt,twoside]{IEEEtran}
\usepackage{graphicx}
\usepackage{amsmath,amssymb}
\usepackage{cases}
\usepackage{color}
\usepackage{amsfonts}
\usepackage{indentfirst}
\usepackage{slashbox}
\usepackage{cite}
\usepackage[justification=centering]{caption}
\usepackage{diagbox}
\usepackage{bm}
\usepackage{amssymb}
\usepackage{caption}
\usepackage{algorithm}
\usepackage{algpseudocode}
\usepackage{booktabs}
\usepackage{subfigure}
\usepackage{multirow}
\usepackage{amsthm}
\usepackage{chngpage}
\usepackage{epstopdf}
\usepackage{hyperref}
\usepackage{pstricks}
\usepackage{blkarray}
\usepackage{arydshln}
\makeatletter
\newtheoremstyle{mythm}{3pt}{3pt}{}{16pt}{\bfseries}{:}{.5em}{}
\theoremstyle{mythm}
\newtheorem{theorem}{Theorem}
\newtheorem{example}{Example}
\newtheorem{definition}{Definition}
\newtheorem{remark}{Remark}

\newtheorem{corollary}{Corollary}
\newtheorem{lemma}{Lemma}
\newtheorem{construction}{Construction}

\newcommand{\tabincell}[2]{\begin{tabular}{@{}#1@{}}#2\end{tabular}}
\begin{document}
\title{PDA Construction via Union of Cartesian Product Cache Configurations for Coded Caching
\author{Jinyu~Wang, Minquan~Cheng,~\IEEEmembership{Member,~IEEE,} Kai~Wan,~\IEEEmembership{Member,~IEEE,}
and~Giuseppe~Caire,~\IEEEmembership{Fellow,~IEEE}
\thanks{J. Wang and M. Cheng are with the Key Lab of Education Blockchain and Intelligent Technology, Ministry of Education, and also with the Guangxi Key Lab of Multi-source Information Mining $\&$ Security, Guangxi Normal University, 541004 Guilin, China (e-mail: mathwjy@163.com, chengqinshi@hotmail.com). J. Wang is also with School of Mathematics and Statistics, Guangxi Normal University, 541004 Guilin, China.}
\thanks{K. Wan is  with the School of Electronic Information and Communications, 
Huazhong University of Science and Technology, 430074  Wuhan, China,  (e-mail: kai\_wan@hust.edu.cn). The work of K.~Wan was partially funded by the   National Natural
Science Foundation of China (NSFC-12141107).}
\thanks{G. Caire is with the Electrical Engineering and Computer Science Department, Technische Universit\"{a}t Berlin,
10587 Berlin, Germany (e-mail: kai.wan@tu-berlin.de, caire@tu-berlin.de).  The work of G.~Caire was partially funded by the European Research Council under the ERC Advanced Grant N. 789190, CARENET.}
}
}

\date{}
\maketitle

\begin{abstract}
Caching is an efficient technique to reduce peak traffic by storing popular content in local caches. Placement delivery array (PDA) proposed by Yan {\it et al.} is a combinatorial structure to design coded caching schemes with uncoded placement and one-shot linear delivery. By taking the $m$-fold Cartesian product of a small base PDA, Wang {\it et al.} constructed a big PDA while maintaining the memory ratio and transmission load unchanged, which achieves linear growth in both the number of users and coded caching gain. In order to achieve exponential growth in both the number of users and coded caching gain, in this paper we propose a PDA construction by taking the union operation of the cache configurations from the $m$-fold Cartesian product of a base PDA. The resulting PDA leads to a coded caching scheme with subpacketization increasing sub-exponentially with the number of users while keeping the load constant for fixed memory ratio. By applying the proposed construction to existing base PDAs, three new coded caching schemes are obtained, which cover some existing schemes as special cases and can achieve lower load with simultaneously lower subpacketization for some memory ratios. 

\end{abstract}

\section{Introduction}
 Caching is well-known technique to reduce traffic congestion by shifting traffic from peak to off-peak hours.
In cache-aided networks, parts of popular content is duplicated into local cache memory during off-peak hours. If the cached content is required during peak hours, it can be retrieved locally, thereby reducing the network traffic, which is referred to as the ``local caching gain''.

Coded caching technique has been widely used to introduce coded multiplexing gain, besides the local caching gain. 
In the seminal work on coded caching by  Maddah-Ali and Niesen (MN)~\cite{MN},   a $(K,M,N)$ caching system was considered, where a server containing $N$ equal-size files connects to $K$ users through an error-free shared link. Each user has its own dedicated cache of size $M$ files. A coded caching scheme operates in two phases: {\em placement} and {\em delivery}. In the placement phase, each file is split into $F$ packets, and some packets are stored in the caches without knowing the users' demands, where $F$ is referred to as the {\em subpacketization}.  In the delivery phase, each user requests a file from the server and the server broadcasts coded messages to all users, such that each user can recover its requested file. The worst case transmission amount (normalized by a file) over all possible demands is defined as the {\em transmission load} or {\em load} denoted by $R$, and 
  $g=K(1-M/N)/R$ is referred to as the {\em coded caching gain}, which represents the average number of users served by each broadcasted message.

By meticulously designing an uncoded placement and a linear coded delivery, the MN scheme \cite{MN} achieved a load which is order optimal within a factor of $2$ with respect to the information-theoretic lower bound \cite{YMA2019}. 
Moreover, when $K\leq N$, the MN scheme was proved to be optimal under uncoded placement~\cite{WTP2016,WTP2020,YMA2018}. However, the subpacketization of the MN scheme increases exponentially with the number of users. In order to reduce the subpacketization, Shanmugam {\it et al.}~\cite{SJTLD} proposed a grouping method, which divides the users into several groups and applies the MN scheme in each group. Yan {\it et al.}~\cite{YCTC} proposed an $F\times K$ array called placement delivery array (PDA) to design coded caching schemes with uncoded placement and one-shot linear delivery (one-shot means that each user can decode a desired packet from at most one transmitted message with the help of its cache). The PDA resulting to the MN scheme is referred to as the MN PDA. Based on the concept of PDA, various coded caching schemes with reduced subpacketization with respect to the MN scheme were proposed in \cite{YCTC,YTCC,CJYT,CJWY,CJTY,ZCW,CWZW,AST,WCLC,WCWC}.
PDA has been extended to different coded caching scenarios like in Device-to-Device (D2D) networks as D2D placement delivery array (DPDA) \cite{WCYT}, in Combination networks as combinational PDA (CPDA) \cite{CLZW}, and in multiple-input single-output (MISO) broadcast channel as multiple-antenna placement delivery array (MAPDA) \cite{YWCQC}. Other combinatorial structures were also used to design coded caching schemes, such as the linear block codes~\cite{TR}, the special $(6,3)$-free hypergraphs \cite{SZG}, the strong edge coloring of bipartite graphs~\cite{YTCC},  and the projective geometry~\cite{CKSM}, etc. The parameters of existing coded caching schemes are summarized in Table \ref{knownnewPDA}. Note that the scheme in \cite{WCLC} is a generalization or improvement of the schemes in \cite{YCTC,CJYT,CWZW,TR,SZG}.

\begin{table}
  \centering
  \caption{Parameters of existing and proposed new shared-link coded caching schemes. Define $\left[k \atop t\right]_p=\frac{(p^k-1)\cdots(p^{k-t+1}-1)}{(p^t-1)\cdots(p-1)}$. \label{knownnewPDA}}
  \begin{tabular}{ccccc}
\toprule
Schemes                          &\tabincell{c}{Number of users} & \tabincell{c}{Memory ratio $M/N$}& \tabincell{c}{Load $R$ }& \tabincell{c}{Subpacketization $F$} \\
\hline
\tabincell{c}{MN scheme in\cite{MN}, $t\in\mathbb{Z}^{+},t\leq K$}    & $K$  & $\frac{t}{K}$ &  $\frac{K-t}{t+1}$ & ${K\choose t}$\\
 \hline

 \tabincell{c}{The grouping method\cite{SJTLD,CJWY},\\ $q,z\in\mathbb{Z}^{+}$, $z\leq q\leq K$} &
 $K$ &  $\frac{z}{q}$   & $\frac{K}{q}\frac{q-z}{z+1}$ & $\frac{q}{\gcd(q,K)} {q\choose z}$\\
\hline

\tabincell{c}{Scheme in \cite{WCLC}, \\$m,q,z,t\in\mathbb{Z}^{+}$,$z<q$, $t\leq m$}
 & ${m\choose t}q^t$ & $1-(\frac{q-z}{q})^t$  & $\frac{(q-z)^t}{\lfloor\frac{q-1}{q-z}\rfloor^t}$& $\lfloor\frac{q-1}{q-z}\rfloor^tq^{m-1}$\\

 \hline

 \tabincell{c}{Scheme in \cite{WCWC}, \\$m,q,z\in \mathbb{Z}^+$,$z<q$} & $mq$& $\frac{z}{q}$ & $\frac{q-z}{z}$ & $z{q\choose z}^m$\\
\hline

\tabincell{c}{Scheme in \cite{YTCC}, $a,b,H\in \mathbb{Z}^{+},$\\ $r\in\mathbb{Z}^{*}$, $r<a<H$,\\$r<b<H$, $a+b\leq H+r$} &  ${H\choose a}$ &  $1-\frac{{a\choose r}{H-a\choose b-r}}{{H\choose b}}$   & \tabincell{c}{$\frac{{H\choose a+b-2r}}{{H\choose b}}\cdot$\\ $\min\{{H-a-b+2r\choose r},$\\ ${{a+b-2r\choose a-r}}\}$} & ${H\choose b}$\\
\hline

\multirow{2}{*}{\tabincell{c}{Scheme in \cite{CKSM},\\ $m,t,k\in \mathbb{Z}^{+}, m+t\leq k$,\\ prime power $p\geq 2$}} & \tabincell{c}{$\frac{1}{t!}p^{\frac{t(t-1)}{2}}$\\$\prod_{i=0}^{t-1}\left[k-i \atop 1\right]_p$}& \tabincell{c}{$1-p^{mt}$\\$\prod_{i=0}^{m-1}\frac{\left[k-t-i \atop 1\right]_p}{\left[k-i \atop 1\right]_p}$} & \tabincell{c}{$\frac{1}{m!}p^{\frac{m(m-1)}{2}}$\\$\prod_{i=0}^{m-1}\left[k-i \atop 1\right]_p$} & \tabincell{c}{$\frac{m!p^{mt}}{(m+t)!}p^{\frac{t(t-1)}{2}}$\\$\prod_{i=0}^{t-1}\left[k-m-i \atop 1\right]_p$}\\  \cdashline{2-5}

 & $\left[k \atop t\right]_p$ & $1-\frac{\left[k-t \atop m\right]_p}{\left[k \atop m+t\right]_p}$ & $\left[k \atop m+t\right]_p$ & $\frac{\left[k \atop m\right]_p}{\left[k \atop m+t\right]_p}$\\
 \hline



\tabincell{c}{Scheme A in Corollary \ref{MNbs},\\$m,t,q,z\in\mathbb{Z}^+$, $q>z$, $t<m$} &
   ${m\choose t}q^{t}$ &  $1-(\frac{q-z}{q})^t$   & $(\frac{q-z}{z})^t$ & $z^t{q\choose z}^{m}$\\
   \hline

\tabincell{c}{Scheme B in Corollary \ref{MNtranbs},\\ $m,t,q,z\in\mathbb{Z}^+$, $q>z$, $t<m$} &
   ${m\choose t}{q\choose z}^t$ &  $1-(\frac{q-z}{q})^t$   & $\frac{{q-1\choose z}^t}{z^t}$ & $z^tq^{m}$\\
   \hline

\tabincell{c}{Scheme C in Corollary \ref{thbsg2},\\ $m,t,q\in\mathbb{Z}^+$, $q\geq 2,t<m$}    & ${m\choose t}q^{2t}$  & $1-(\frac{q-1}{q})^t$ &  $\frac{(q-1)^tq^t}{2^t}$ & $2^mq^m$\\

 \bottomrule

\end{tabular}

\end{table}

\subsection{Contributions}
The grouping method in \cite{SJTLD} directly replicates a given $K_1$-user PDA $m$ times to obtain an $mK_1$-user PDA, which implies that each user selects one from the $K_1$ cache configurations determined by the original $K_1$-user PDA as its cache state. This can reduce the subpacketization significantly with respect to the MN scheme, but the load will increase linearly with $m$, since the coded caching gain does not change. In order to reduce the subpacketization while maintaining the load as low as possible, our recent work~\cite{WCWC}  constructed an $mK_1$-user PDA by taking the $m$-fold Cartesian product of a $K_1$-user base PDA (which is a PDA satisfying some constraints, please refer to Definition \ref{def-basePDA} for detail), while keeping the memory ratio and load unchanged. This implies that both the number of users ($mK_1$) and the coded caching gain ($mg_1$) increase linearly with $m$.
In this paper, we aim to achieve exponential growth in both the number of users and the coded caching gain simultaneously.  Precisely, given a $K_1$-user base PDA with coded caching gain $g_1$, we aim to construct a PDA with number of users ${m\choose t}K_1^t$ and coded caching gain ${m\choose t}g_1^t$, where $1\leq t\leq m$. To the best of our knowledge, none of existing PDA constructions have achieved this goal. The main difficulties lie in two aspects. The first one is how to design the placement of ${m\choose t}K_1^t$ users based on the placement of $K_1$ users from the base PDA. Direct replication is the simplest method, but it cannot achieve synchronous growth of coded caching gain, while the $m$-fold Cartesian product of the base PDA can generate $mK_1$ different cache configurations, which ensures synchronous linear growth of the coded caching gain.  In this paper, by dividing the $mK_1$ cache configurations into $m$ groups and taking the union of any $t$ cache configurations from $t$ different groups, we obtain ${m\choose t}K_1^t$ different cache state for ${m\choose t}K_1^t$  users. 
The second one is how to ahcieve coded caching gain ${m\choose t}g_1^t$ based on the well designed placement. Based on the special structure of the union of Cartesian product cache configurations, we divide the ${m\choose t}K_1^t$ users into ${m\choose t}$ groups, and achieve a multicast gain of $g_1^t$ within each group. Meanwhile by combining the constraints of base PDA, we take the XOR of the multicast messages for different groups to obtain final multicast messages, each of which is useful for ${m\choose t}g_1^t$ users, thus achieving a total coded caching gain of ${m\choose t}g_1^t$.

More precisely, the main contributions of this paper are as follows.

$\bullet$ Given a $K_1$-user PDA $\mathbf{P}$ with subpacketization $F_1$, memory ratio $\frac{Z_1}{F_1}$ and coded caching gain $g_1$, if it is a base PDA with parameter $\lambda$, for any positive integers $m$ and $t$ with $t\leq m$,  we construct an ${m\choose t}K_1^t$-user PDA $\mathbf{P}_{m,t}$, leading to a $(K,M,N)$ coded caching scheme with memory ratio $1-(\frac{F_1-Z_1}{F_1})^t$, coded caching gain ${m\choose t}g_1^t$, and subpacketization $\lambda^{t-m}F_1^m$, which increases sub-exponentially with the number of users when $t>1$. It covers the PDA construction in \cite{WCWC} as a special case when $t=1$. 

$\bullet$ As applications of the above PDA construction, three new coded caching schemes (i.e., Scheme A, B, C) are obtained, which are listed in the last three rows of Table \ref{knownnewPDA}. Scheme A covers the schemes in \cite[Theorem 3]{WCWC} and \cite[Theorem 18]{SZG} as special cases, while   Schemes B, C covers the schemes in \cite[Theorem 4]{WCWC} and \cite[Theorem 5]{WCWC} as special cases, respectively. Compared to the scheme in \cite{WCLC}, which is a generalization or improvement of the schemes in \cite{YCTC,CJYT,CWZW,TR,SZG}, Scheme A achieves lower load with higher subpacketization, while Schemes B, C can achieve lower load with simultaneously lower subpacketization. Compared to the schemes in \cite{SJTLD,CJWY,CKSM}, Scheme A,B,C can achieve lower load with simultaneously lower subpacketization for some memory ratios.

The rest of this paper is organized as follows. In Section \ref{prem}, the coded caching system, the definition of PDA, and some related works are introduced. In Section \ref{mainresult}, we list main results of this paper, including the main result of the PDA construction and the obtained three new coded caching schemes. In Section \ref{sec_method}, the main idea and general steps of the PDA construction are provided.  Performance analysis of the new schemes is given in Section \ref{performance}.  Finally, Section \ref{conclusion} concludes the paper while a proof is provided in Appendix \ref{prPDAPmt}.

{\bf Notations:}
$|\cdot|$ denotes the cardinality of a set. For any integers $a,b$ with $a\leq b$, $[a:b]:=\left\{ a,a+1,\ldots,b\right\}$.
If $a$ is not divisible by $q$, $\langle a\rangle_q$ denotes the least non-negative residue of $a$ modulo $q$; otherwise, $\langle a\rangle_q:=q$.
For any positive integers $m$ and $t$ with $t\leq m$, let ${[1:m]\choose t}=\{\mathcal{T}\ |\   \mathcal{T}\subseteq [1:m], |\mathcal{T}|=t\}$, i.e., ${[1:m]\choose t}$ is the collection of all $t$-size subsets of $[1:m]$.
Let $\mathcal{B}=\{b_1,b_2,\ldots,b_{n}\}$ be a set of numbers with $b_1<b_2<\ldots<b_n$, for any $i\in[1:n]$, $\mathcal{B}[i]$ denotes the $i$-th smallest element of $\mathcal{B}$, i.e., $\mathcal{B}[i]=b_i$.
For any $m\times n$ array $\mathbf{P}$, $\mathbf{P}(i,j)$ represents the element located in the $i$-th row and the $j$-th column of $\mathbf{P}$, $\mathbf{P}^\top$ denotes the transpose array of $\mathbf{P}$.

\section{System model and related works}
\label{prem}
In this section, we will introduce the coded caching system proposed in \cite{MN}, the definition of PDA \cite{YCTC} and related works in \cite{WCWC}. 
\subsection{System Model}
In the coded caching system proposed by Maddah-Ali and Niesen \cite{MN}, a server containing $N$ equal-size files $\mathcal{W}=\{W_1, \ldots, W_N\}$ connects to $K$ users $U_1,\ldots,U_K$ through an error-free shared link where $K\leq N$. Each user has a cache of size $M$ files ($0\leq M\leq N$). A $(K,M,N)$ coded caching scheme consists of two phases:

$\bullet$ {\bf Placement phase:} Each file $W_n$ is divided equally into $F$ packets, i.e., $W_{n}=\{W_{n,j}|j\in[1:F]\}$. The server populates each user's cache with some packets (or coded packets). If packets are cached directly without coding, it is called uncoded placement. Otherwise, it is called coded placement. The contents cached by user $U_k$ are denoted by $\mathcal{Z}_{U_k}$. The placement phase is done without knowledge of the user demands.

$\bullet$ {\bf Delivery phase:} Each user requests one file from the server independently. The request vector is denoted by $\mathbf{d}=(d_1,\cdots,d_{K})$, which means that user $U_k$ requests the file $W_{d_k}$, where $d_k\in[1:N]$ and $k\in[1:K]$. Then the server broadcasts coded packets of size $R_{{\bf d}}$ files to the users such that each user can recover its desired file with the help of its cached contents.

The worst case transmission load (or load) of a coded caching scheme is defined by 
$R=\max_{{\bf d}\in[1:N]^K }R_{\bf d}$,
that is, the maximum number of files transmitted by the server over all possible user demands.

\subsection{Placement Delivery Array}
\label{PDA}
For the coded caching system, Yan et al. \cite{YCTC} proposed a combinatorial structure called placement delivery array (PDA) to design coded caching schemes with uncoded placement and one-shot linear delivery.

\begin{definition}(\cite{YCTC})
\label{def-PDA}
For  positive integers $K,F, Z$ and $S$, an $F\times K$ array  $\mathbf{P}$ 
composed of a specific symbol $``*"$  and $S$ integers in $[1:S]$, is called a $(K,F,Z,S)$ placement delivery array (PDA) if it satisfies the following conditions:

{\bf C1:} The symbol $``*"$ appears $Z$ times in each column;

{\bf C2:} Each integer in $[1:S]$ occurs at least once in the array;

{\bf C3:} For any two distinct entries $\mathbf{P}(j_1,k_1)$ and $\mathbf{P}(j_2,k_2)$, if    $\mathbf{P}(j_1,k_1)=\mathbf{P}(j_2,k_2)=s\in[1:S]$, then $\mathbf{P}(j_1,k_2)=\mathbf{P}(j_2,k_1)=*$, i.e., the corresponding $2\times 2$  subarray formed by rows $j_1,j_2$ and columns $k_1,k_2$ must be of the following form
    \begin{eqnarray*}
    \left(\begin{array}{cc}
      s & *\\
      * & s
    \end{array}\right)~\textrm{or}~
    \left(\begin{array}{cc}
      * & s\\
      s & *
    \end{array}\right).
  \end{eqnarray*}
\end{definition}
If each integer appears exactly $g$ times in $\mathbf{P}$, $\mathbf{P}$ is called a $g$-regular $(K,F,Z,S)$ PDA, $g$-$(K,F,Z,S)$ PDA or $g$-PDA for short.

A $(K,F,Z,S)$ PDA $\mathbf{P}$ is an $F\times K$ array composed of a specific symbol $``*"$ and $S$ integers, where the $j$-th row represents the $j$-th packet of all files and the $k$-th column represents user $U_k$. If $\mathbf{P}(j,k)=*$, it means the $j$-th packet of all files is cached by user $U_k$. If $\mathbf{P}(j,k)=s$ is an integer, it represents that the $j$-th packet of all files is not cached by user $U_k$, then the XOR of all the requested packets indicated by $s$ is broadcasted by the server at time slot $s$. Condition C1 of Definition \ref{def-PDA} implies that memory ratio of each user is $\frac{Z}{F}$.
Condition C3 of Definition \ref{def-PDA} ensures that each user can recover its desired packet, since all the other packets in the broadcasted message are cached by it. Condition C2 of Definition \ref{def-PDA} implies that the number of messages broadcasted by the server is exactly $S$, so the load is $R=\frac{S}{F}$.


\begin{lemma}(\cite{YCTC})
\label{th-Fundamental}
If there exists a $(K,F,Z,S)$ PDA, there always exists a $(K,M,N)$ coded caching scheme with the memory ratio $\frac{M}{N}=\frac{Z}{F}$, subpacketization $F$ and load $R=\frac{S}{F}$.
\end{lemma}

\subsection{PDA construction via Cartesian product}
\label{Con-PDA}
Next we will briefly introduce the PDA construction via Cartesian product proposed in \cite{WCWC}, which constructs an $mK_1$-user PDA by taking the $m$-fold Cartesian product of a special $K_1$-user PDA (called base PDA), while keeping the memory ratio and load unchanged. The Cartesian product of two arrays is defined as follows.
\begin{definition}({\it Cartesian product}, \cite{WCWC})
\label{defcar}
Let $\mathbf{A}=({\bf a}_1^{\top},{\bf a}_2^{\top}, \ldots, {\bf a}_{F_1}^{\top})^{\top}$ be an $F_1\times K_1$ array,
 and $\mathbf{B}=({\bf b}_1^{\top},{\bf b}_2^{\top}, \ldots, {\bf b}_{F_2}^{\top})^{\top}$ be an $F_2\times K_2$ array. The Cartesian product of $\mathbf{A}$ and $\mathbf{B}$   is an $F_1F_2\times (K_1+K_2)$ array,  defined by \eqref{def_car_prod}.
\begin{figure}
\begin{equation}
\label{def_car_prod}
\small
\mathbf{A}\times \mathbf{B}=\left(\begin{array}{cc}
{\bf a}_1 & {\bf b}_1\\
\vdots         &  \vdots\\
{\bf a}_{F_1} & {\bf b}_1\\
\vdots &\vdots \\
{\bf a}_{1} & {\bf b}_{F_2} \\
\vdots         &  \vdots\\
{\bf a}_{F_1} & {\bf b}_{F_2}\\
\end{array}\right).
\end{equation}
\vspace{-1.2cm}
\end{figure}
In particular, for any positive integer $m$, the $m$-fold Cartesian product of $\mathbf{A}$ is an $F_1^m\times mK_1$ array, defined by
$\mathbf{A}^{m}=\underbrace{\mathbf{A}\times \mathbf{A}\times\ldots\times\mathbf{A}}_{m}$.
\end{definition}
A base PDA is a PDA satisfying some constraints, defined as follows.
\begin{definition}({\it Base PDA}, \cite{WCWC})
\label{def-basePDA}
A $(K_1,F_1,Z_1,S_1)$ PDA $\mathbf{P}$ is called a base PDA with parameter $\lambda$, if it satisfies the following conditions:

{\bf C4:} There exists a positive integer $\lambda$ dividing both $F_1$ and $Z_1$, such that $\mathbf{P}(j,k)=*$ if and only if $\mathbf{P}(\langle j\rangle _{\frac{F_1}{\lambda}},k)=*$ for any $j\in[1:F_1]$ and $k\in[1:K_1]$;

{\bf C5:} There exists a mapping $\phi$ from $[1:S_1]$ to $[1:\frac{F_1}{\lambda}]$ such that
      \begin{itemize}
      \item[--] for each integer $s\in[1:S_1]$, for any $\mathbf{P}(j,k)=s$ where $j\in[1:F_1]$ and $k\in[1:K_1]$, the $\phi(s)$-th row of $\mathbf{P}$ is a {\it star row} for  $s$;
      \item[--] for each   $j\in [1:\frac{F_1}{\lambda}]$, by defining $\mathcal{B}_j =\{s|\phi(s)=j, s\in[1:S_1]\}$, it holds that  $|\mathcal{B}_j|=\frac{\lambda S_1}{F_1}$.
      \end{itemize}
\end{definition}

Wherein, the definition of star row is as follows.
\begin{definition}({\it Star row},\cite{WCWC})
\label{def-starrow}
For a $(K_1,F_1,Z_1,S_1)$ PDA $\mathbf{P}$ and an integer $s\in[1:S_1]$, we say that the $i$-th row of $\mathbf{P}$ is a star row for $s$ if each column  of  $\mathbf{P}$ containing $s$  has  $*$ in  the    $i$-th row, i.e., we have $\mathbf{P}(i,k)=*$ for any $\mathbf{P}(j,k)=s$ where $j\in[1:F_1]$ and $k\in[1:K_1]$.
\end{definition}

Condition C4 means that the $F_1\times K_1$ PDA $\mathbf{P}$ can be divided into $\lambda$ subarrays, i.e., $\mathbf{P}=(\mathbf{A}_{1}^{\top},\ldots,\mathbf{A}_{\lambda}^{\top})^{\top}$, where each subarray $\mathbf{A}_{i}$ has the same dimension (i.e., $\frac{F_1}{\lambda}\times K_1$) and the same placement of stars (i.e., $\mathbf{P}(j,k)=*$ if and only if $\mathbf{P}(\langle j\rangle _{\frac{F_1}{\lambda}},k)=*$). Obviously, any PDA satisfies Condition C4 with parameter $\lambda=1$. Condition C5 implies that there is at least one star row for each integer in $\mathbf{P}$. Moreover, there exists a mapping $\phi$ for assigning exactly one star row in the first $\frac{F_1}{\lambda}$ rows for each integer, such that each of the first $\frac{F_1}{\lambda}$ rows is assigned to the same number of integers (i.e., $\lambda S_1/F_1$ integers). In short, Condition C5 means that there exists a uniform partition of $[1:S_1]$, i.e., $\mathcal{B}_1, \mathcal{B}_2, \ldots, \mathcal{B}_{\frac{F_1}{\lambda}}$, where $\mathcal{B}_j$ denotes the set of integers whose assigned star row is the $j$-th row of $\mathbf{P}$. For example, the following $2$-$(4,4,2,4)$ PDA
\begin{eqnarray}
\label{eq_basePDA}
\mathbf{P}=\left(\begin{array}{cccc}
*&*&3&1\\
2&*&*&4\\
1&3&*&*\\
*&2&4&*\\
\end{array}\right)
\end{eqnarray}
is a base PDA with parameter $\lambda=1$. First, $\mathbf{P}$ satisfies Condition C4. Second, $1$ appears twice in $\mathbf{P}$, i.e., $\mathbf{P}(3,1)=\mathbf{P}(1,4)=1$, since $\mathbf{P}(4,1)=\mathbf{P}(4,4)=*$, the forth row is a star row for integer $1$. Similarly, the first, second and third rows are star rows for $2, 3, 4$ respectively. Let $\phi(1)=4$, $\phi(2)=1$, $\phi(3)=2$ and $\phi(4)=3$, we have $\mathcal{B}_1=\{2\}$, $\mathcal{B}_2=\{3\}$, $\mathcal{B}_3=\{4\}$ and $\mathcal{B}_4=\{1\}$, then $|\mathcal{B}_j|=1$ for any $j\in[1:4]$. $\mathbf{P}$ satisfies Condition C5, so it is a base PDA with parameter $\lambda=1$.

Some existing PDAs are not base PDA. However, any $g$-PDA with the same number of stars in each row can be transformed into a base PDA. For example, the $2$-$(2,2,1,1)$ MN PDA $\mathbf{Q}$ in \eqref{eq_extran} is not a base PDA, since there is no star row for integer $1$. It can be transformed into the $1$-$(2,2,1,2)$ base PDA with parameter $\lambda=1$ denoted by $\mathbf{P}$ in \eqref{eq_extran}.
\begin{eqnarray}
\label{eq_extran}
\mathbf{Q}=\left(\begin{array}{cc}
*&1\\
1&*\\
\end{array}\right) \ \ \rightarrow \ \ 
\mathbf{P}=\left(\begin{array}{cc}
	*&2\\
	1&*\\
\end{array}\right).
\end{eqnarray}
\begin{lemma}(\cite{WCWC})
\label{letransform}
If there exists a $g_1$-$(K_1,F_1,Z_1,S_1)$ PDA with the same number of stars in each row, where $g_1\geq 2$, there always exists a $(g_1-1)$-$(K_1,(g_1-1)F_1,(g_1-1)Z_1,g_1S_1)$ base PDA with parameter $\lambda=g_1-1$.
\end{lemma}

It is worth noting that any existing PDA has the same number of stars in each row \cite{WCWC}, so any existing PDA can be transformed into a base PDA. In addition, the authors constructed a new base PDA as follows.
\begin{lemma}(\cite{WCWC})
\label{ex_basePDA}
For any positive integer $q\geq 2$, there exists a $2$-$(q^2,2q,2,(q-1)q^2)$ base PDA with parameter $\lambda=1$.
\end{lemma}
Based on a base PDA, the PDA construction via Cartesian product is summarized as follows, and an example is provided in Section \ref{sec_method} to illustrate its main steps.
\begin{lemma}(\cite{WCWC})
\label{lemma-pm}
If there exists a $g_1$-$(K_1,F_1,Z_1,S_1)$ base PDA with parameter $\lambda$, then for any positive integer $m$, there always exists an $mg_1$-$\left(mK_1,\lambda (\frac{F_1}{\lambda})^m, Z_1(\frac{F_1}{\lambda})^{m-1},\right.$ $\left.S_1(\frac{F_1}{\lambda})^{m-1}\right)$ PDA $\mathbf{P}_m$, which leads to a $(K, M, N)$ coded caching scheme with number of users $K=mK_1$, user memory ratio $\frac{M}{N}=\frac{Z_1}{F_1}$, subpacketization $F=\lambda (\frac{F_1}{\lambda})^m$ and load $R=\frac{S_1}{F_1}$.
\end{lemma}

\section{Main results}
\label{mainresult}
In this paper, we propose a PDA construction by taking the union of the Cartesian product cache configurations, which can achieve exponential growth in both the number of users and the coded caching gain simultaneously. Specifically, we first take the $m$-fold Cartesian product of a $K_1$-user base PDA to obtain $mK_1$ cache configurations, which are uniformly divided into $m$ groups. Then we take the union of any $t$ cache configurations from $t$ different groups as a user's cache state, thus the placement array composed of ``*" and null entries is obtained. Finally, we design the null entries in the placement array meticulously, such that the resulting array is a PDA with the coded caching gain as large as possible. The main result of this construction is as follows.

\begin{theorem}
\label{thmulpmt}
If there exists a $g_1$-$(K_1, F_1, Z_1, S_1)$ base PDA with parameter $\lambda$, then for any positive integers $m,t$ with $t\leq m$, there always exists an ${m\choose t}g_1^t$-$\left({m\choose t}K_1^t, \lambda^t (\frac{F_1}{\lambda})^m,\lambda^t (\frac{F_1}{\lambda})^m\cdot\right.$ $\big.(1-(\frac{F_1-Z_1}{F_1})^t),(\frac{F_1}{\lambda})^{m-t}S_1^t\big)$ PDA $\mathbf{P}_{m,t}$, which leads to a $(K,M,N)$ coded caching scheme with the number of users $K={m\choose t}K_1^t$, memory ratio $\frac{M}{N}=1-\big(\frac{F_1-Z_1}{F_1}\big)^t$, subpacketization $F=\lambda^t (\frac{F_1}{\lambda})^m$, and load $R=\big(\frac{S_1}{F_1}\big)^t$.
\end{theorem}
The construction of the PDA in Theorem \ref{thmulpmt} is given in Section \ref{sec_method} and the detailed proof is provided in Appendix \ref{prPDAPmt}.

Since the MN PDA is a $(z+1)$-$\left(q,{q\choose z},{q-1\choose z-1},{q\choose z+1}\right)$ PDA with $z$ stars in each row \cite{YCTC}, it can be transformed into a $z$-$\left(q,z{q\choose z},z{q-1\choose z-1},(z+1){q\choose z+1}\right)$ base PDA with parameter $\lambda=z$ from Lemma \ref{letransform},
then the following new coded caching scheme can be obtained from Theorem \ref{thmulpmt}.
\begin{corollary}({\bf Scheme A})
\label{MNbs}
For any positive integers $q,z,m,t$ with $z<q$ and $t\leq m$, there exists an ${m\choose t} z^t$-$\left({m\choose t}q^t, z^t{q\choose z}^{m}, z^t{q\choose z}^{m}(1-(\frac{q-z}{q})^t), (q-z)^t{q\choose z}^{m}\right)$ PDA, which leads to a $(K,M,N)$ coded caching scheme with the number of users $K={m\choose t}q^t$, memory ratio $\frac{M}{N}=1-\big(\frac{q-z}{q}\big)^t$, subpacketization $F=z^t{q\choose z}^{m}$, and load $R=(\frac{q-z}{z})^t$.
\end{corollary}

Since the transpose MN PDA is a $(z+1)$-$\left({q\choose z},q,z,{q\choose z+1}\right)$ PDA with ${q-1\choose z-1}$ stars in each row \cite{CJTY}, it can be transformed into a $z$-$\left({q\choose z},qz,z^2,(z+1){q\choose z+1}\right)$ base PDA with parameter $\lambda=z$, then the following new coded caching scheme can be obtained from Theorem \ref{thmulpmt}.
\begin{corollary}({\bf Scheme B})
\label{MNtranbs}
For any positive integers $q,z,m,t$ with $z<q$ and $t\leq m$, there exists an ${m\choose t} z^t$-$\left({m\choose t}{q\choose z}^t, z^tq^{m}, z^tq^{m}(1-(\frac{q-z}{q})^t), q^{m}{q-1\choose z}^{t}\right)$ PDA, which leads to an $(K,M,N)$ coded caching scheme with the number of users $K={m\choose t}{q\choose z}^t$, memory ratio $\frac{M}{N}=1-\big(\frac{q-z}{q}\big)^t$, subpacketization $F=z^tq^{m}$, and load $R=\frac{{q-1\choose z}^t}{z^t}$.
\end{corollary}

By applying Theorem \ref{thmulpmt} to the base PDA in Lemma \ref{ex_basePDA}, the following new coded caching scheme can be obtained.
\begin{corollary}({\bf Scheme C})
\label{thbsg2}
For any positive integers $q,m,t$ with $q\geq 2$ and $t\leq m$, there exists an ${m\choose t}2^t$-$\left({m\choose t}q^{2t},2^{m}q^{m},2^{m}q^{m}(1-(\frac{q-1}{q})^t),2^{m-t}(q-1)^tq^{m+t}\right)$ PDA, which leads to a $(K,M,N)$ coded caching scheme with the number of users $K={m\choose t}q^{2t}$, memory ratio $\frac{M}{N}=1-\big(\frac{q-1}{q}\big)^t$, subpacketization $F=(2q)^{m}$, and load $R=\frac{(q-1)^tq^t}{2^t}$.
\end{corollary}
\begin{remark}
\label{re_share}
1) When $t=1$, Theorem \ref{thmulpmt} reduces to Lemma \ref{lemma-pm}; the schemes in Corollary \ref{MNbs}, \ref{MNtranbs}, \ref{thbsg2} are exactly the schemes in \cite[Theorem 3,4,5]{WCWC} respectively. When $z=1$, the schemes in Corollary \ref{MNbs}, \ref{MNtranbs} have the same parameters as the scheme in \cite[Theorem 18]{SZG}.

2) The number of users in Corollary \ref{MNbs}, \ref{MNtranbs}, \ref{thbsg2} is $K={m\choose t}K_1^t$ where $K_1=q,{q
    \choose z}, q^2$ respectively. Since $(\frac{m}{t})^t<{m\choose t}<(\frac{em}{t})^t$, we have $\frac{t}{eK_1}K^{\frac{1}{t}}<m<\frac{t}{K_1}K^{\frac{1}{t}}$. The subpacketization of the schemes in Corollary \ref{MNbs}, \ref{MNtranbs}, \ref{thbsg2} are $F_{\text{A}}=z^t{q\choose z}^m$, $F_{\text{B}}=z^tq^m$, $F_{\text{C}}=2^mq^m$ respectively, which all increase sub-exponentially with the number of users $K$ when $m$ increases and $t>1$.

3) Since the load of the schemes in Corollary \ref{MNbs}, \ref{MNtranbs}, \ref{thbsg2} is independent of $m$, when the number of users $K$ is between ${m \choose t}K_1^t$ and ${m+1 \choose t}K_1^t$, we can use the ${m+1 \choose t}K_1^t$-user PDA by regarding the missing users as virtual users, which achieves the same load as the ${m \choose t}K_1^t$-user PDA, while the subpacketization is at most a constant multiple of that of the ${m \choose t}K_1^t$-user PDA.
\end{remark}

\section{The PDA construction in Theorem \ref{thmulpmt}}
\label{sec_method}
In this section, we first use an example to introduce the steps of the PDA construction in \cite{WCWC}, then illustrate the main idea of the PDA construction in Theorem \ref{thmulpmt} and provide the mathematical representation.

\begin{example}
\label{example2}
As mentioned earlier, the array $\mathbf{P}$ in \eqref{eq_basePDA} is a $g_1$-$(K_1,F_1,Z_1,S_1)$ base PDA with parameter $\lambda=1$, where $K_1=F_1=4$, $Z_1=2$, $S_1=4$ and $g_1=2$. When $m=3$, an $mK_1=12$-user PDA $\mathbf{P}_m=\mathbf{P}_3$ can be constructed by the PDA construction proposed in \cite{WCWC}, which contains two steps, as illustrated in Fig. \ref{fig1-1}. In the first step, the $m=3$-fold Cartesian product of the base PDA $\mathbf{P}$ is taken to obtain an $F_1^m\times mK_1=64\times 12$ cache configuration array $\mathbf{C}$, which satisfies
  \begin{equation}
  \label{productarray}
  \mathbf{C}((f_1,f_2,f_3),(\delta,b))=\mathbf{P}(f_\delta,b).
  \end{equation}
In the second step, the integers in $\mathbf{C}$ are adjusted to satisfy Condition C3 of Definition \ref{def-PDA}, such that the resulting array denoted by $\mathbf{P}_3$ is an $mK_1=12$-user PDA with coded caching gain $mg_1=6$. Obviously, the number of users and coded caching gain increase linearly with $m$.

\begin{figure}
  \centering
  \includegraphics[width=6.5in]{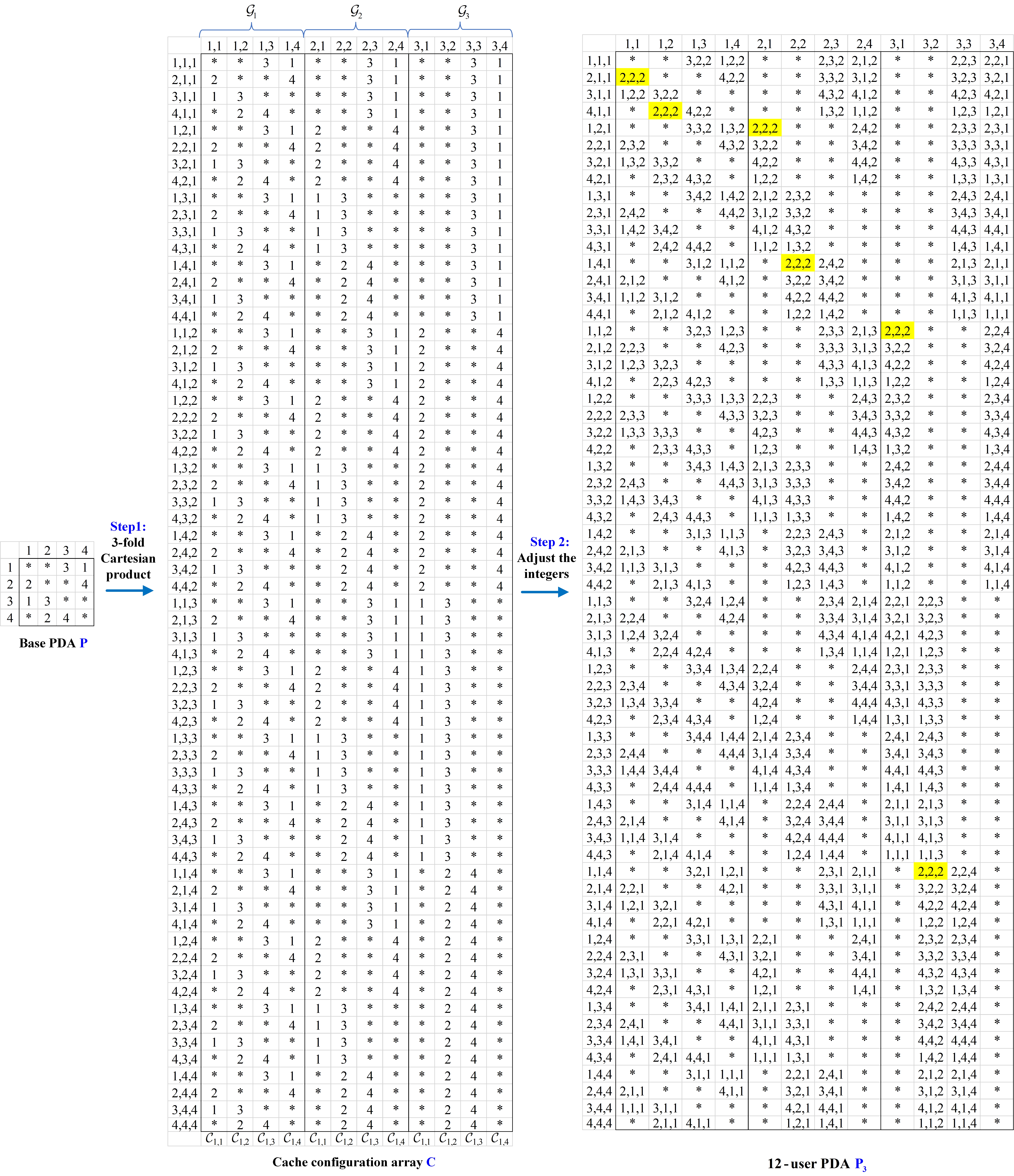}\\
  \caption{The steps of generating the $12$-user PDA $\mathbf{P}_{3}$ from the base PDA $\mathbf{P}$ in \cite{WCWC}.}\label{fig1-1}
\end{figure}

In order to achieve exponential growth in both the number of users and the coded caching gain simultaneously, we introduce a new parameter $t$ where $t\in[1:m]$. Our goal is to achieve the number of users $K={m\choose t}K_1^t$ and the coded caching gain $g={m\choose t}g_1^t$. For example, when $m=3,t=2$, we will construct an ${m\choose t}K_1^t=48$-user PDA $\mathbf{P}_{m,t}=\mathbf{P}_{3,2}$ with coded caching gain ${m\choose t}g_1^t=12$ based on the same base PDA $\mathbf{P}$ by the following two steps.

{\bf Step 1. Placement array.} We first take the $m=3$-fold Cartesian Product of the base PDA $\mathbf{P}$ to obtain the cache configuration array $\mathbf{C}$ as done in \cite{WCWC} (see Fig. \ref{fig1-1}), which contains $12$ cache configurations, i.e., $$\mathcal{C}_{\delta,b}=\{(f_1,f_2,f_3)|\mathbf{C}((f_1,f_2,f_3),(\delta,b))=*,f_i\in[1:F_1]=[1:4],i\in[1:3]\}$$ where $\delta\in[1:m]=[1:3]$ and $b\in[1:K_1]=[1:4]$. Each $\mathcal{C}_{\delta,b}$ represents a caching strategy, where the packet indexed by $(f_1,f_2,f_3)$ is cached if $(f_1,f_2,f_3)\in \mathcal{C}_{\delta,b}$. From Fig. \ref{fig1-1}, the $12$ cache configurations are as follows:
\begin{align*}
\mathcal{C}_{1,1}&=\{(1,j,k),(4,j,k)|j,k\in[1:4]\}, \ \mathcal{C}_{1,2}=\{(1,j,k),(2,j,k)|j,k\in[1:4]\},\\
\mathcal{C}_{1,3}&=\{(2,j,k),(3,j,k)|j,k\in[1:4]\}, \ \mathcal{C}_{1,4}=\{(3,j,k),(4,j,k)|j,k\in[1:4]\};\\
\mathcal{C}_{2,1}&=\{(i,1,k),(i,4,k)|i,k\in[1:4]\}, \ \mathcal{C}_{2,2}=\{(i,1,k),(i,2,k)|i,k\in[1:4]\},\\
\mathcal{C}_{2,3}&=\{(i,2,k),(i,3,k)|i,k\in[1:4]\}, \ \mathcal{C}_{2,4}=\{(i,3,k),(i,4,k)|i,k\in[1:4]\};\\
\mathcal{C}_{3,1}&=\{(i,j,1),(i,j,4)|i,j\in[1:4]\}, \ \ \mathcal{C}_{3,2}=\{(i,j,1),(i,j,2)|i,j\in[1:4]\},\\
\mathcal{C}_{3,3}&=\{(i,j,2),(i,j,3)|i,j\in[1:4]\}, \ \ \mathcal{C}_{3,4}=\{(i,j,3),(i,j,4)|i,j\in[1:4]\}.
\end{align*}
The cache configurations are divided into $m=3$ groups, i.e., $\mathcal{G}_1=\{\mathcal{C}_{1,1},\mathcal{C}_{1,2},\mathcal{C}_{1,3},\mathcal{C}_{1,4}\}$,  $\mathcal{G}_2=\{\mathcal{C}_{2,1},\mathcal{C}_{2,2},\mathcal{C}_{2,3},\mathcal{C}_{2,4}\}$ and  $\mathcal{G}_3=\{\mathcal{C}_{3,1},\mathcal{C}_{3,2},\mathcal{C}_{3,3},\mathcal{C}_{3,4}\}$.

Then we take the union of any $t=2$ cache configurations from $t=2$ different groups to generate $K={m\choose t}K_1^t=48$ cache states, which are represented by the placement array $$\mathbf{P}_{m,t}^*=\mathbf{P}_{3,2}^*=(\mathbf{G}_{1,2} \ \mathbf{G}_{1,3} \ \mathbf{G}_{2,3}),$$
where each column represents a cache state and $\mathbf{G}_{i,j}$ includes the columns obtained by taking the union of $2$ cache configurations from group $\mathcal{G}_{i}$ and group $\mathcal{G}_{j}$. Due to page size limitations, we only show the first subarray $\mathbf{G}_{1,2}$ in Fig. \ref{fig2}.
Note that each column of the placement array $\mathbf{P}_{3,2}^*$ is indexed by $(\{\delta_1,\delta_2\},(b_1,b_2))$   (where $\delta_1,\delta_2\in[1:m]=[1:3]$ with $\delta_1<\delta_2$ and $b_1,b_2\in[1:K_1]=[1:4]$), which implies that the corresponding cache state is the union of the $b_1$-th cache configuration of group $\mathcal{G}_{\delta_1}$ and the $b_2$-th cache configuration of group $\mathcal{G}_{\delta_2}$, i.e., $\mathcal{C}_{\delta_1,b_1}\cup \mathcal{C}_{\delta_2,b_2}$.
  Since there exists a one-to-one relationship between the cache configurations and the columns of $\mathbf{C}$, i.e., the cache configuration $\mathcal{C}_{\delta,b}$ is corresponding to the column of $\mathbf{C}$ indexed by $(\delta,b)$, for ease of expression, we refer to the union of cache configurations $\mathcal{C}_{\delta_1,b_1}$ and $\mathcal{C}_{\delta_2,b_2}$ as the union of the columns indexed by $(\delta_1,b_1)$ and $(\delta_2,b_2)$. The resulting column indexed by $(\{\delta_1,\delta_2\},(b_1,b_2))$ will contain a null entry in a particular row if and only if both the entries in the same row of the columns indexed by $(\delta_1,b_1)$ and $(\delta_2,b_2)$ are integers.
  For example, the first column of $\mathbf{P}_{3,2}^*$ indexed by $(\{1,2\},(1,1))$ is the union of the columns of $\mathbf{C}$ indexed by $(1,1)$ and $(2,1)$, please refer to Fig. \ref{fig1-1} and Fig. \ref{fig2}.
  \begin{figure}
  \centering
  \includegraphics[width=6.5in]{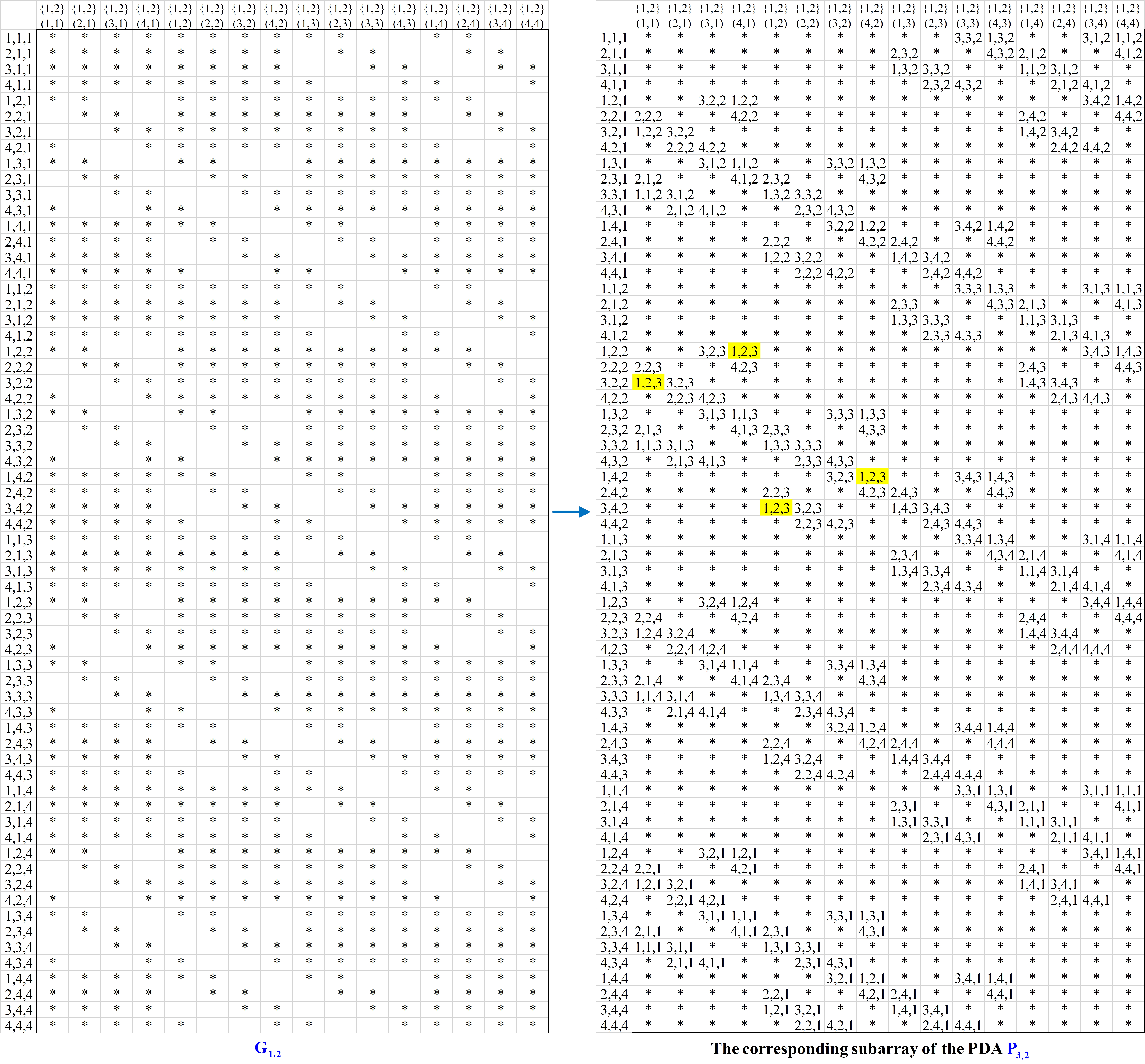}\\
  \caption{The first subarray of the placement array $\mathbf{P}_{3,2}^*$ and the corresponding subarray of the PDA $\mathbf{P}_{3,2}$.}\label{fig2}
  \vspace{-0.8cm}
  \end{figure}

{\bf Step 2. Delivery array.}
In this step we fill each null entry in the placement array $\mathbf{P}_{m,t}^{*}$ with an $m$-dimensional vector $\mathbf{e}$ belonging to
\begin{equation}
\label{eq_S}
\mathcal{S}=\{(e_1,\ldots,e_m)|e_i=\mathcal{B}_{r_i}[\nu_i],r_i\in[1:\frac{F_1}{\lambda}],\nu_i\in[1:\frac{\lambda S_1}{F_1}],i\in[1:m],\nu_1=\ldots=\nu_{m-t+1}\},
\end{equation}
such that the resulting array denoted by $\mathbf{P}_{m,t}$ contains $S=|\mathcal{S}|=(\frac{F_1}{\lambda})^{m-t}S_1^t$ different vectors and satisfies Condition C3 of Definition \ref{def-PDA}.
Note that $\mathcal{B}_1,\ldots,\mathcal{B}_{\frac{F_1}{\lambda}}$ is a uniform partition of $[1:S_1]$, $\mathcal{B}_{r_i}$ represents the set of integers whose assigned star row is the $r_i$-th row of the base PDA $\mathbf{P}$, $\mathcal{B}_{r_i}[\nu_i]$ represents the $\nu_i$-th smallest element of $\mathcal{B}_{r_i}$.
For each vector $\mathbf{e}\in\mathcal{S}$, we will fill vector $\mathbf{e}$ into the placement array $\mathbf{P}_{m,t}^*$ in the following way.
For any $t$-size subset of $[1:m]$, say $\mathcal{T}=\{\delta_1,\ldots,\delta_t\}$ where $\delta_1<\ldots<\delta_t$, assume that $\mathcal{T}\cap[1:m-t+1]=\{\delta_1,\ldots,\delta_w\}$, then $|[m-t+2:m]\setminus \mathcal{T}|=w-1$, assume that $[m-t+2:m]\setminus \mathcal{T}=\{\sigma_1,\ldots,\sigma_{w-1}\}$ where $\sigma_1<\ldots<\sigma_{w-1}$. For any $i\in[1:t]$, let
\begin{equation}
\label{e-s}
s_{\delta_i}^{\mathcal{T}}=\begin{cases} \mathcal{B}_{r_{\delta_i}}[\nu_{\delta_i}]\ \ \  \text{if} \ i\in\{1\}\cup[w+1:t],\\
\mathcal{B}_{r_{\delta_i}}[\nu_{\sigma_{i-1}}] \ \ \  \text{if} \ i\in[2:w].
\end{cases}
\end{equation}
Let the subarray of $\mathbf{P}$ containing $s_{\delta_i}^{\mathcal{T}}$ be $\mathbf{P}^{(s_{\delta_i}^{\mathcal{T}})}$, since each integer appears $g_1$ times in $\mathbf{P}$, the dimension of $\mathbf{P}^{(s_{\delta_i}^{\mathcal{T}})}$ is $g_1\times g_1$. Assume that the Cartesian product of all these subarrays is
\begin{equation}
\label{ed_Ps}
\mathbf{P}^{(s_{\delta_1}^{\mathcal{T}})}\times\ldots\times\mathbf{P}^{(s_{\delta_t}^{\mathcal{T}})}=(\mathbf{Q}_{\delta_1}^{\mathcal{T}} \ldots \mathbf{Q}_{\delta_t}^{\mathcal{T}}),
\end{equation}
where each subarray $\mathbf{Q}_{\delta_i}^{\mathcal{T}}$ has dimension $g_1^t\times g_1$.
Let
\begin{equation}
\label{eq_QT}
\mathbf{Q}^{\mathcal{T}}=(\mathbf{Q}_{1}^{\mathcal{T}} \ldots \mathbf{Q}_{m}^{\mathcal{T}}),
\end{equation}
where each $\mathbf{Q}_{i}^{\mathcal{T}}$ is a $g_1^t\times g_1$ array. If $i\in\mathcal{T}$, $\mathbf{Q}_{i}^{\mathcal{T}}$ is defined by \eqref{ed_Ps}; if $i\notin \mathcal{T}$, each entry of $\mathbf{Q}_{i}^{\mathcal{T}}$ is a star.
Assume that all the $t$-size subsets of $[1:m]$ are $\mathcal{T}_1, \ldots, \mathcal{T}_{{m\choose t}}$, let
\begin{equation}
\label{eq_Q}
\mathbf{Q}=\left(\begin{array}{c}
\mathbf{Q}^{\mathcal{T}_1}\\
\vdots\\
\mathbf{Q}^{\mathcal{T}_{{m\choose t}}}
\end{array}
\right)=(\mathbf{D}_1 \ldots \mathbf{D}_m),
\end{equation}
where each subarray $\mathbf{D}_i$ has dimension ${m\choose t}g_1^t\times g_1$ for $i\in[1:m]$.
By taking the union of any $t$ columns of $\mathbf{Q}$ (where each column belongs to different $\mathbf{D}_i$), an ${m\choose t}g_1^t\times {m\choose t}g_1^t$ subarray of the placement array $\mathbf{P}_{m,t}^{*}$ can be obtained, which is equivalent to the following array
\begin{equation*}
\mathbf{P}_{m,t}^{(\mathbf{e})}=\left(\begin{array}{cccc}
\Box &*& \cdots& *\\
*&\Box&\cdots&*\\
\vdots&\vdots& \ddots & \vdots\\
*&*&\cdots&\Box
\end{array}
\right)
\end{equation*}
with respect to row/column permutation. In $\mathbf{P}_{m,t}^{(\mathbf{e})}$, only the main diagonal elements are null entries (denoted by the symbol $``\Box"$), the other elements are all stars. Fill each null entry in $\mathbf{P}_{m,t}^{(\mathbf{e})}$ by the vector $\mathbf{e}$, then $\mathbf{e}$ appears ${m\choose t}g_1^t$ times in the resulting array $\mathbf{P}_{m,t}$ which satisfies Condition C3 of Definition \ref{def-PDA}.

In this example, the base PDA in \eqref{eq_basePDA} can be written as
\begin{equation}
\label{eq_basePDA1}
\mathbf{P}=\bordermatrix{
 &1 & 2& 3 & 4\cr
1&*  & * &3_2&1_2\cr
2&2_1& * &*  &4_2\cr
3&1_1&3_1&*  &*  \cr
4&*  &2_2&4_1&*
},
\end{equation}where the two integer $s$'s are written as $s_1$ and $s_2$ from left to right for $s\in[1:4]$.
Since $m=3$, $t=2$, $F_1=S_1=4$, $\lambda=1$, $\mathcal{B}_1=\{2\}$, $\mathcal{B}_2=\{3\}$, $\mathcal{B}_3=\{4\}$ and $\mathcal{B}_4=\{1\}$, we have $\mathcal{S}=\{(e_1,e_2,e_3)|e_i\in[1:4],i\in[1:3]\}$ from \eqref{eq_S}. For any $s\in[1:S_1]=[1:4]$, the subarray of $\mathbf{P}$ containing $s$ is denoted by $\mathbf{P}^{(s)}$, then we have
\begin{equation}
\label{eq_subarray_s}
\mathbf{P}^{(1)}=\left(\begin{array}{cc}
 * &1_2  \\
1_1&* \\
\end{array}\right),
\mathbf{P}^{(2)}=\left(\begin{array}{cc}
2_1&*\\
* &2_2\\
\end{array}\right),
\mathbf{P}^{(3)}=\left(\begin{array}{cc}
 * &3_2\\
3_1&*\\
\end{array}\right),
\mathbf{P}^{(4)}=\left(\begin{array}{cc}
 * &4_2\\
4_1&*\\
\end{array}\right)
\end{equation}from \eqref{eq_basePDA1}.

For any vector $\mathbf{e}\in \mathcal{S}$, without loss of generality, assume that $\mathbf{e}=(1,2,3)$, we have $e_1=1=\mathcal{B}_4[1]$, $e_2=2=\mathcal{B}_1[1]$ and $e_3=3=\mathcal{B}_2[1]$, i.e., $r_1=4, r_2=1, r_3=2$ and $\nu_1=\nu_2=\nu_3=1$ from \eqref{eq_S}. For $t=2$-size subset $\mathcal{T}_1=\{1,2\}$, we have $s_1^{\mathcal{T}_1}=1, s_2^{\mathcal{T}_1}=2$ from \eqref{e-s}, then we have
$$\mathbf{P}^{(s_{1}^{\mathcal{T}_1})}\times \mathbf{P}^{(s_{2}^{\mathcal{T}_1})}=\mathbf{P}^{(1)}\times \mathbf{P}^{(2)}=\left(\begin{array}{cc|cc} *& 1_2 & 2_1 & *\\ 1_1 & * &2_1 &*\\ *& 1_2 & * & 2_2\\ 1_1 & *& * & 2_2 \end{array}\right)=(\mathbf{Q}_{1}^{\mathcal{T}_1} \ \mathbf{Q}_{2}^{\mathcal{T}_1}),$$
$$\mathbf{Q}^{\mathcal{T}_1}=(\mathbf{Q}_{1}^{\mathcal{T}_1} \ \mathbf{Q}_{2}^{\mathcal{T}_1} \ \mathbf{Q}_{3}^{\mathcal{T}_1})=\left(\begin{array}{cc|cc|cc} *& 1_2 & 2_1 & *& * & *\\ 1_1 & * &2_1 &*& * & *\\ *& 1_2 & * & 2_2& * & *\\ 1_1 & *& * & 2_2& * & * \end{array}\right)$$
from \eqref{ed_Ps}, \eqref{eq_QT} and \eqref{eq_subarray_s}.
Similarly, for $t=2$-size subset $\mathcal{T}_2=\{1,3\}$, we have $s_{1}^{\mathcal{T}_2}=1$ and $s_{3}^{\mathcal{T}_2}=3$, which leads to
$$\mathbf{P}^{(s_{1}^{\mathcal{T}_2})}\times \mathbf{P}^{(s_{3}^{\mathcal{T}_2})}=\mathbf{P}^{(1)}\times \mathbf{P}^{(3)}=\left(\begin{array}{cc|cc} *& 1_2 & * & 3_2\\ 1_1 & * & * & 3_2\\ *& 1_2 & 3_1 & *\\ 1_1 & *& 3_1 & * \end{array}\right)=(\mathbf{Q}_{1}^{\mathcal{T}_2} \ \mathbf{Q}_{3}^{\mathcal{T}_2}),$$
$$\mathbf{Q}^{\mathcal{T}_2}=(\mathbf{Q}_{1}^{\mathcal{T}_2} \ \mathbf{Q}_{2}^{\mathcal{T}_2} \ \mathbf{Q}_{3}^{\mathcal{T}_2})=\left(\begin{array}{cc|cc|cc} *& 1_2 &*&*& * & 3_2\\ 1_1 & * &*&*& * & 3_2\\ *& 1_2 &*&*& 3_1 & *\\ 1_1 & *&*&*& 3_1 & * \end{array}\right).$$

For $t=2$-size subset $\mathcal{T}_3=\{2,3\}$, we have $s_{2}^{\mathcal{T}_3}=2$ and $s_{3}^{\mathcal{T}_3}=3$, which leads to
$$\mathbf{P}^{(s_{2}^{\mathcal{T}_3})}\times \mathbf{P}^{(s_{3}^{\mathcal{T}_3})}=\mathbf{P}^{(2)}\times \mathbf{P}^{(3)}=\left(\begin{array}{cc|cc} 2_1& * & * & 3_2\\ * & 2_2 & * & 3_2\\ 2_1& * & 3_1 & *\\ * & 2_2& 3_1 & * \end{array}\right)=(\mathbf{Q}_{2}^{\mathcal{T}_3} \ \mathbf{Q}_{3}^{\mathcal{T}_3}),$$
$$\mathbf{Q}^{\mathcal{T}_3}=(\mathbf{Q}_{1}^{\mathcal{T}_3} \ \mathbf{Q}_{2}^{\mathcal{T}_3} \ \mathbf{Q}_{3}^{\mathcal{T}_3})=\left(\begin{array}{cc|cc|cc} *& * &2_1& * & * & 3_2\\ *& * &* & 2_2 & * & 3_2\\ *& * &2_1& * & 3_1 & *\\ *& * &* & 2_2& 3_1 & * \end{array}\right).$$
Consequently we have
$$\mathbf{Q}=\left(\begin{array}{c} \mathbf{Q}^{\mathcal{T}_1}\\ \mathbf{Q}^{\mathcal{T}_2}\\ \mathbf{Q}^{\mathcal{T}_3} \end{array} \right)=
\left(\begin{array}{cc|cc|cc}
*& 1_2 & 2_1 & *& * & *\\
1_1 & * &2_1 &*& * & *\\
 *& 1_2 & * & 2_2& * & *\\
 1_1 & *& * & 2_2& * & *\\
 *& 1_2 &*&*& * & 3_2\\
 1_1 & * &*&*& * & 3_2\\
 *& 1_2 &*&*& 3_1 & *\\
 1_1 & *&*&*& 3_1 & *\\
 *& * &2_1& * & * & 3_2\\
 *& * &* & 2_2 & * & 3_2\\
  *& * &2_1& * & 3_1 & *\\
  *& * &* & 2_2& 3_1 & *
\end{array}\right)
=(\mathbf{D}_1 \ \mathbf{D}_2 \ \mathbf{D}_3 )$$ from \eqref{eq_Q}.
By taking the union of any $t=2$ columns of $\mathbf{Q}$ (each column from different $\mathbf{D}_i$), the $12\times 12$ subarray of $\mathbf{P}_{3,2}^{*}$ denoted by $\mathbf{P}_{3,2}^{(\mathbf{e})}$ in \eqref{eq_P32e} can be obtained,
\begin{figure}
\begin{small}
\begin{eqnarray}
\label{eq_P32e}
\mathbf{P}_{3,2}^{(\mathbf{e})}=(\mathbf{D}_{1,2} \ \mathbf{D}_{1,3} \ \mathbf{D}_{2,3})= \ \ \ \ \ \ \ \ \ \ \ \ \ \ \ \ \ \ \ \ \ \ \ \ \ \ \ \ \ \ \ \ \ \ \ \ \ \ \ \ \ \ \ \ \ \ \ \ \ \ \ \ \ \ \ \ \ \ \ \ \ \ \ \ \ \ \ \ \ \ \ \ \ \ \ \ \ \ \ \ \ \ \ \ \ \ \notag\\
\begin{blockarray}{ccccccccccccc}
     &\{1,2\}\atop(1,1)&\{1,2\}\atop(4,1)&\{1,2\}\atop(1,2)&\{1,2\}\atop(4,2)&\{1,3\}\atop(1,2)&\{1,3\}\atop(4,2)&\{1,3\}\atop(1,3)&\{1,3\}\atop(4,3)&\{2,3\}\atop(1,2)&\{2,3\}\atop(2,2)&\{2,3\}\atop(1,3)&\{2,3\}\atop(2,3)\\
     \begin{block}{c(cccc|cccc|cccc)}
(1,2,2)&*&\square&  *  & * & * & * & *&  *  & * & * & * & * \\
(3,2,2)&\square&*&  *  & * & * & * & *&  *  & * & * & * & * \\
(1,4,2)& *  & * & * & \square & * & *&*&  *  & * & * & * & * \\
(3,4,2)& *&  *  &\square & * & *&* & * &*&  *  & * & * & *\\
(1,1,1)&*&*&  *  & * & * & * & *& \square & * & * & * & * \\
(3,1,1)&*&*&  *  & * & * & * &\square&  *  & * & * & * & * \\
(1,1,3)& *  & * & * & * & * & \square&*&  *  & * & * & * & * \\
(3,1,3)& *&  *  & *& * & \square&* & * &*&  *  & * & * & *\\
(4,2,1)&*&*&  *  & * & * & * & *& * & * & * & \square & * \\
(4,4,1)&*&*&  *  & * & * & * &*&  *  & * & * & * & \square \\
(4,2,3)& *  & * & * & * & * & *&*&  *  & \square & * & * & * \\
(4,4,3)& *&  *  & *& * & *&* & * &*&  *  & \square & * & *\\
\end{block}
\end{blockarray}
\end{eqnarray}
\end{small}
\vspace{-1.2cm}
\end{figure}
where $\mathbf{D}_{i,j}$ is obtained by taking the union of two columns from $\mathbf{D}_i$ and $\mathbf{D}_j$. Note that the first column of $\mathbf{D}_{i,j}$ is the union of the first column of $\mathbf{D}_i$ and the first column of $\mathbf{D}_j$; the second column of $\mathbf{D}_{i,j}$ is the union of the second column of $\mathbf{D}_i$ and the first column of $\mathbf{D}_j$; the third column of $\mathbf{D}_{i,j}$ is the union of the first column of $\mathbf{D}_i$ and the second column of $\mathbf{D}_j$; the forth column of $\mathbf{D}_{i,j}$ is the union of the second column of $\mathbf{D}_i$ and the second column of $\mathbf{D}_j$. It can be seen that there is exactly one null entry in each row or column of $\mathbf{P}_{3,2}^{(\mathbf{e})}$, the row and column indexes of each null entry are determined as follows. For example, the first column of $\mathbf{P}_{3,2}^{(\mathbf{e})}$ is the union of the first and third columns of $\mathbf{Q}$, since $\mathbf{Q}(2,1)=1_1$ and $\mathbf{Q}(2,3)=2_1$, the entry at the second row of the first column of $\mathbf{P}_{3,2}^{(\mathbf{e})}$ is a null entry, whose row index is $(3,2,2)$ and column index is $(\{1,2\},(1,1))$. In the column index, the set $\{1,2\}$ represents the corresponding column is the union of two columns from $\mathbf{D}_1$ and $\mathbf{D}_2$; the first coordinate of the vector $(1,1)$ is the column index of $\mathbf{Q}(2,1)=1_1$ in $\mathbf{P}$, i.e., $1$ from \eqref{eq_basePDA1}; the second coordinate of the vector $(1,1)$ is the column index of $\mathbf{Q}(2,3)=2_1$ in $\mathbf{P}$, i.e., $1$ from \eqref{eq_basePDA1}. 
In the row index, since the set in the column index is $\{1,2\}=\mathcal{T}_1$, the first coordinate is the row index of $\mathbf{Q}(2,1)=1_1$ in $\mathbf{P}$, i.e., $3$ from \eqref{eq_basePDA1}; the second coordinate is the row index of $\mathbf{Q}(2,3)=2_1$ in $\mathbf{P}$, i.e., $2$ from \eqref{eq_basePDA1}; the third coordinate is the assigned star row for integers $s_3^{\mathcal{T}_2}=3$ and $s_3^{\mathcal{T}_3}=3$, i.e., $2$, since $\mathcal{B}_2=\{3\}$ (which means the set of integers whose assigned star row is the second row is $\{3\}$). The row and column indexes of other null entries in $\mathbf{P}_{3,2}^{(\mathbf{e})}$ can be obtained similarly.
Fill each null entry in $\mathbf{P}_{3,2}^{(\mathbf{e})}$ by the vector $\mathbf{e}$, then $\mathbf{e}$ appears $12$ times and satisfies Condition C3 of Definition \ref{def-PDA}.  The other vectors in $\mathcal{S}$ can be filled into the placement array similarly. Fig. \ref{fig2} shows the subarray of $\mathbf{P}_{3,2}$ corresponding to $\mathbf{G}_{1,2}$.



\end{example}

\begin{figure}
  \centering
  \includegraphics[width=6in]{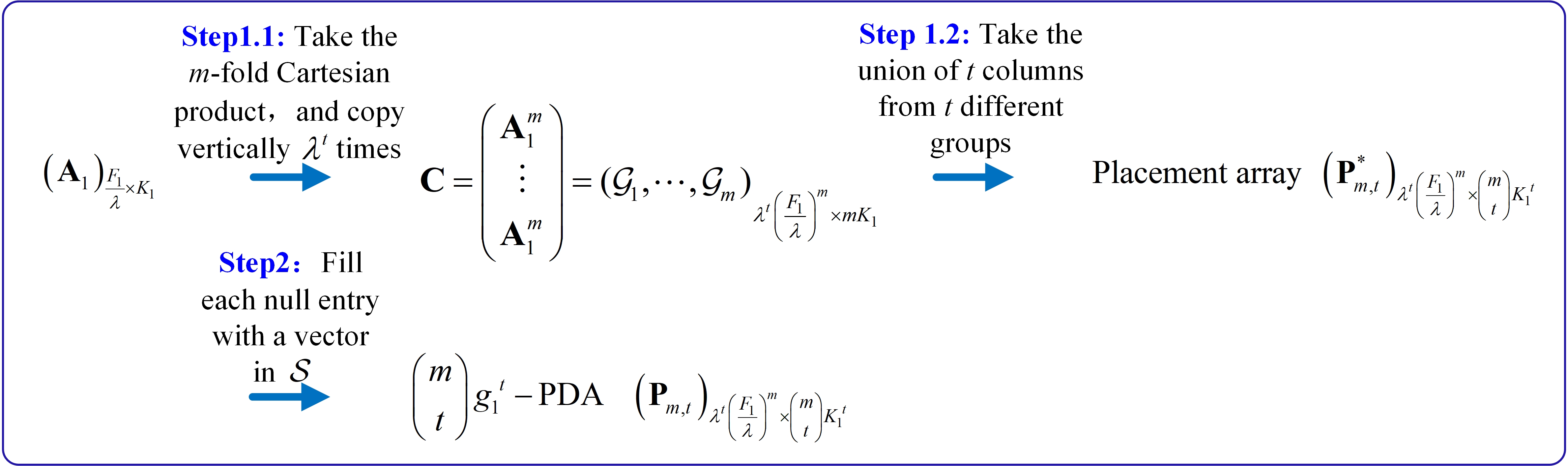}\\
  \caption{The general steps of generating a PDA $\mathbf{P}_{m,t}$ from a base PDA $\mathbf{P}$, where $\mathbf{A}_1$ is the first subarray of the base PDA.}\label{idea}
  \vspace{-1.1cm}
\end{figure}

In general, given a $g_1$-$(K_1,F_1,Z_1,S_1)$ base PDA $\mathbf{P}$ with parameter $\lambda$, from Condition C4 of Definition \ref{def-basePDA}, $\mathbf{P}$ can be divided into $\lambda$ subarrays, i.e., $\mathbf{P}=(\mathbf{A}_{1}^{\top},\ldots,\mathbf{A}_{\lambda}^{\top})^{\top}$,
where each subarray $\mathbf{A}_{i}$ has the same dimension (i.e., $\frac{F_1}{\lambda}\times K_1$) and the same placement of stars. The main idea of the construction is shown in Fig. \ref{idea}. We first take the $m$-fold Cartesian product of the first subarray $\mathbf{A}_1$ and copy it $\lambda^t$ times vertically to obtain the $\lambda^t(\frac{F_1}{\lambda})^m\times mK_1$ cache configuration array $\mathbf{C}$, whose rows are indexed by $(\mathbf{f},\bm{\varepsilon})$ where $\mathbf{f}=(f_1,\ldots,f_m)\in[1:\frac{F_1}{\lambda}]^m$ and $\bm{\varepsilon}=(\varepsilon_1,\ldots,\varepsilon_t)\in[1:\lambda]^t$, columns are indexed by $(\delta,b)$ where $\delta\in[1:m]$ and $b\in[1:K_1]$, satisfying
\begin{equation}
\label{eq_C}
\mathbf{C}((\mathbf{f},\bm{\varepsilon}),(\delta,b))=\mathbf{P}(f_\delta,b).
\end{equation}
Each column of $\mathbf{C}$ corresponds to a cache configuration, and the $mK_1$ cache configurations are uniformly divided into $m$ groups, denoted by $\mathcal{G}_1,\ldots,\mathcal{G}_m$ from left to right. It is worth noting that when $\lambda=1$, we only need to take the $m$-fold Cartesian product of the entire base PDA $\mathbf{P}$ since $\mathbf{P}=\mathbf{A}_1$, and the vector $\bm{\varepsilon}$ in the row index of the cache configuration array $\mathbf{C}$ can be omitted since in this case $\bm{\varepsilon}$ is always $(1,\ldots,1)$, as done in Example \ref{example2}. Then we take the union of $t$ columns from $t$ different groups to generate ${m\choose t}K_1^t$ cache states, which are represented by the placement array $\mathbf{P}_{m,t}^*$. Finally, by filling each null entry of $\mathbf{P}_{m,t}^*$ by a vector in the set $\mathcal{S}$ defined in \eqref{eq_S} as illustrated in Example \ref{example2}, an ${m\choose t}K_1^t$-user PDA $\mathbf{P}_{m,t}$ with coded caching gain ${m\choose t}g_1^t$ can be obtained.

The mathematical representation of the above PDA construction for Theorem \ref{thmulpmt} is as follows. Note that $\mathcal{B}_1,\mathcal{B}_2,\ldots,\mathcal{B}_{\frac{F_1}{\lambda}}$ is a uniform partition of $[1:S_1]$, where $\mathcal{B}_j$ denotes the set of integers whose assigned star row is the $j$-th row of $\mathbf{P}$; $\mathcal{B}_l[\mu]$ denotes the $\mu$-th smallest element of the set $\mathcal{B}_l$.
\begin{construction}
\label{constr1}
Given a $g_1$-$(K_1,F_1,Z_1,S_1)$ base PDA $\mathbf{P}$ with parameter $\lambda$, for any positive integers $m,t$ with $t\leq m$, let the row index set be $\mathcal{F}=[1:\frac{F_1}{\lambda}]^m\times[1:\lambda]^t$ and column index set be $\mathcal{K}={[1:m]\choose t}\times [1:K_1]^t$. For any $({\bf f},\bm{\varepsilon})\in\mathcal{F}$ and $(\mathcal{T},{\bf b})\in \mathcal{K}$ where ${\bf f}=(f_1,\ldots,f_{m})\in[1:\frac{F_1}{\lambda}]^m$, $\bm{\varepsilon}=(\varepsilon_1,\ldots,\varepsilon_{t})\in[1:\lambda]^t$, $\mathcal{T}=\{\delta_1,\ldots, \delta_{t}\}\in{[1:m]\choose t}$ with $\delta_1<\ldots<\delta_{t}$ and ${\bf b}=(b_1,\ldots,b_{t})\in [1:K_1]^t$, let $\mathcal{T}\cap [1:m-t+1]=\{\delta_1,\ldots,\delta_{w}\}$, $[m-t+2:m]\setminus \mathcal{T}=\{\sigma_1,\ldots,\sigma_{w-1}\}$ with $\sigma_1<\ldots<\sigma_{w-1}$. A $\lambda^t(\frac{F_1}{\lambda})^{m}\times {m\choose t}K_1^t$ array $\mathbf{P}_{m,t}=(\mathbf{P}_{m,t}(({\bf f},\bm{\varepsilon}),(\mathcal{T},{\bf b})))_{({\bf f},\bm{\varepsilon})\in\mathcal{F},(\mathcal{T},{\bf b})\in\mathcal{K}}$ is defined as follows
\begin {equation}
\label{constrPm}
\mathbf{P}_{m,t}(({\bf f},\bm{\varepsilon}),(\mathcal{T},{\bf b}))=\begin{cases}
{\bf e} &\text{if} \ \ \mathbf{P}(f_{\delta_h},b_h)\neq *  \ \text{for any} \ \ h\in[1:t],\\
 *    & \text{otherwise}, \\
\end{cases}
\end{equation}
where ${\bf e}=(e_1,e_2,\ldots,e_{m})$ satisfying
\begin {equation}
\label{constre}
e_i=\begin{cases}
\mathcal{B}_{l_h}[\mu_1] &\text{if} \ \ i=\delta_h, h\in[1:w],\\
\mathcal{B}_{l_h}[\mu_h] & \text{if} \ \ i=\delta_h, h\in[w+1:t],\\
\mathcal{B}_{f_i}[\mu_{h+1}] & \text{if} \ \ i=\sigma_h, h\in[1:w-1], \\
\mathcal{B}_{f_i}[\mu_1] & \text{otherwise}, \\
\end{cases}
\end{equation}
where
 $\mathbf{P}(f_{\delta_h}+(\varepsilon_h-1)\frac{F_1}{\lambda},b_h)=\mathcal{B}_{l_h}[\mu_h]$\footnote{From Condition C4 of Definition \ref{def-basePDA}, $\mathbf{P}(f_{\delta_h},b_h)\neq *$ implies $\mathbf{P}(f_{\delta_h}+(\varepsilon_h-1)\frac{F_1}{\lambda},b_h)\in[1:S_1]$. Since  $\mathcal{B}_1,\mathcal{B}_2,\ldots,\mathcal{B}_{\frac{F_1}{\lambda}}$ is a uniform partition of $[1:S_1]$, there exist unique $l_h$ and $\mu_h$ such that $\mathbf{P}(f_{\delta_h}+(\varepsilon_h-1)\frac{F_1}{\lambda},b_h)=\mathcal{B}_{l_h}[\mu_h]$.} for any $h\in[1:t]$.
\end{construction}

When $m=3,t=2$, based on the base PDA $\mathbf{P}$ in \eqref{eq_basePDA} with $\lambda=1$, $K_1=F_1=4$, $\mathcal{B}_1=\{2\}$, $\mathcal{B}_2=\{3\}$, $\mathcal{B}_3=\{4\}$ and $\mathcal{B}_4=\{1\}$, the array $\mathbf{P}_{3,2}$ generated by Construction \ref{constr1} is exactly the array $\mathbf{P}_{3,2}$ constructed in Example \ref{example2}, where the vector ${\bm \varepsilon}$ in the row index is omitted, since in each row index ${\bm \varepsilon}$ is $(1,1)$.
For example, when ${\bf f}=(3,2,2)$ and $(\mathcal{T},{\bf b})=(\{1,2\},(1,1))$, we have $f_1=3,f_2=f_3=2$, $\delta_1=1,\delta_2=2$ and $b_1=b_2=1$, then $\mathbf{P}(f_{\delta_1},b_1)=\mathbf{P}(3,1)=1=\mathcal{B}_4[1],\mathbf{P}(f_{\delta_2},b_2)=\mathbf{P}(2,1)=2=\mathcal{B}_1[1]$, so $l_1=4,l_2=1$ and $\mu_1=\mu_2=1$. Since $\mathcal{T}\cap [1:m-t+1]=\{1,2\}$, we have $w=2$ and $[m-t+2:m]\setminus \mathcal{T}=\{3\}=\{\sigma_1\}$. Hence, we have $\mathbf{P}_{3,2}({\bf f},(\mathcal{T},{\bf b}))=(\mathcal{B}_{l_1}[\mu_1],\mathcal{B}_{l_2}[\mu_1],\mathcal{B}_{f_3}[\mu_2])=(1,2,3)$ from \eqref{constre}, which coincides with the subarray of $\mathbf{P}_{3,2}$ in Fig. \ref{fig2}.

In fact, the array generated by Construction \ref{constr1} is an ${m\choose t}g_1^t$-$\big({m\choose t}K_1^t, \lambda^t (\frac{F_1}{\lambda})^m,\lambda^t (\frac{F_1}{\lambda})^m(1-(\frac{F_1-Z_1}{F_1})^t),\big.$ $\left.(\frac{F_1}{\lambda})^{m-t}S_1^t\right)$ PDA, which leads to a $(K,M,N)$ coded caching scheme with the number of users $K={m\choose t}K_1^t$, memory ratio $\frac{M}{N}=1-(\frac{F_1-Z_1}{F_1})^t$, subpacketization $F=\lambda^t (\frac{F_1}{\lambda})^m$, and load $R=(\frac{S_1}{F_1})^t$. For the detailed proof, please refer to Appendix \ref{prPDAPmt}.

\section{Performance Analysis}
\label{performance}

\subsection{Comparison to Existing Shared-link Coded Caching Schemes}
In this section, we will compare the new schemes in Corollary \ref{MNbs}, \ref{MNtranbs}, \ref{thbsg2} with existing schemes listed in Table \ref{knownnewPDA}. Since the new schemes cover the schemes in \cite{SZG,WCWC} as special cases from Remark \ref{re_share}, we will compare the new schemes with the schemes in \cite{MN,SJTLD,CJWY,YTCC,WCLC,CKSM}.



{\it 1) Analytical comparison to the WCLC scheme in \cite{WCLC}}

When $K={m\choose t}q^t$ and $\frac{M}{N}=1-(\frac{q-z}{q})^t$, the subpacketization and load of Scheme A in Corollary \ref{MNbs} are $F_{\text{A}}=z^t{q\choose z}^{m}$ and
$R_{\text{A}}=\left(\frac{q-z}{z}\right)^t$ respectively. The WCLC scheme in \cite{WCLC} achieves subpacketization $F_{\text{WCLC}}=\lfloor\frac{q-1}{q-z}\rfloor^tq^{m-1}$ and load $R_{\text{WCLC}}=\frac{(q-1)^t}{\lfloor\frac{q-1}{q-z}\rfloor^t}$.
Since $(\frac{m}{t})^t<{m\choose t}<(\frac{em}{t})^t$ and $K={m\choose t}q^t$, we have $\frac{t}{eq}K^{\frac{1}{t}}<m<\frac{t}{q}K^{\frac{1}{t}}$, i.e., $m=\Theta(K^{\frac{1}{t}})$ for fixed $t$ and $q$.
Consequently, we have          $$\frac{F_{\text{A}}}{F_{\text{WCLC}}}=\frac{qz^t}{\lfloor\frac{q-1}{q-z}\rfloor^t}\left(\frac{{q\choose z}}{q}\right)^{\Theta(K^{\frac{1}{t}})},\ \ \ \
\frac{R_{\text{A}}}{R_{\text{WCLC}}}=\frac{1}{\big(\frac{z}{\lfloor\frac{q-1}{q-z}\rfloor}\big)^t}.$$
If $z=1$ or $q-1$, Scheme A achieves the same load as the scheme in \cite{WCLC} while the subpacketization is $q$ times that of the scheme in \cite{WCLC}.
If $1<z<q-1$, we have $\frac{q-1}{q-z}<z$, which implies $R_{\text{A}}<R_{\text{WCLC}}$.
Particularly, if $1<z\leq\frac{q}{2}$ and $t>1$, we have $\lfloor\frac{q-1}{q-z}\rfloor=1$, then the load of Scheme A is only $\frac{1}{z^t}$ that of the scheme in \cite{WCLC}, while the subpacketization of Scheme A has a sub-exponential growth compared to the scheme in \cite{WCLC}.

When $K={m\choose t}{q\choose z}^t$ and $\frac{M}{N}=1-(\frac{q-z}{q})^t$, the subpacketization and load of Scheme B in Corollary \ref{MNtranbs} are $F_{\text{B}}=z^tq^{m}$ and $R_{\text{B}}=\frac{{q-1\choose z}^t}{z^t}$ respectively.
By letting $m'=m$, $t'=t$, $q'={q\choose z}$ and $z'={q-1 \choose z-1}$, the subpacketization and load of the scheme in \cite{WCLC} are $F_{\text{WCLC}}=\lfloor\frac{q'-1}{q'-z'}\rfloor^t q'^{m-1}=\big\lfloor\frac{{q\choose z}-1}{{q-1\choose z}}\big\rfloor^t{q \choose z}^{m-1}$ and       $R_{\text{WCLC}}=\big(\frac{q'-z'}{\lfloor\frac{q'-1}{q'-z'}\rfloor}\big)^t=\frac{{q-1\choose z}^t}{\big\lfloor\frac{{q\choose z}-1}{{q-1\choose z}}\big\rfloor^t}$ respectively.
Since $(\frac{m}{t})^t<{m\choose t}<(\frac{em}{t})^t$ and $K={m\choose t}{q\choose z}^t$, we have $\frac{t}{e{q\choose z}}K^{\frac{1}{t}}<m<\frac{t}{{q\choose z}}K^{\frac{1}{t}}$, i.e., $m=\Theta(K^{\frac{1}{t}})$ for fixed $q,z$ and $t$.
Consequently, we have
$$\frac{F_{\text{B}}}{F_{\text{WCLC}}}=\frac{z^t{q \choose z}}{\left\lfloor\frac{{q\choose z}-1}{{q-1\choose z}}\right\rfloor^t}\frac{1}{\left(\frac{{q\choose z}}{q}\right)^{\Theta(K^{\frac{1}{t}})}},\ \ \ \
\frac{R_{\text{B}}}{R_{\text{WCLC}}}=\frac{\left\lfloor\frac{{q\choose z}-1}{{q-1\choose z}}\right\rfloor^t}{z^t}.$$
If $z=1$ or $q-1$, Scheme B achieves the same load as the scheme in \cite{WCLC} while the subpacketization is $q$ times that of the scheme in \cite{WCLC}.
If $1<z<q-1$, leading to $\frac{q}{q-z}\leq z$, then we have $\frac{{q\choose z}-1}{{q-1\choose z}}<\frac{{q\choose z}}{{q-1\choose z}}=\frac{q}{q-z}\leq z$, which implies  $R_{\text{B}}<R_{\text{WCLC}}$. Moreover, the subpacketization of Scheme B has an exponential (if $t=1$) or sub-exponential (if $t>1$) decrease compared to the scheme in \cite{WCLC} when $m$ is large.


When $K={m\choose t}q^{2t}$ and $\frac{M}{N}=1-(\frac{q-1}{q})^t$, the subpacketization and load of Scheme C in Corollary \ref{thbsg2} are $R_{\text{C}}=\frac{(q-1)^tq^t}{2^t}$ and $F_{\text{C}}=2^{m}q^{m}$ respectively. By letting $m'=m$, $t'=t$, $q'=q^2$ and $z'=q$, we have $\lfloor\frac{q'-1}{q'-z'}\rfloor=\lfloor\frac{q^2-1}{q^2-q}\rfloor=1$, then the subpacketization and load of the scheme in \cite{WCLC} are $F_{\text{WCLC}}=\big\lfloor\frac{q'-1}{q'-z'}\big\rfloor^t q'^{m-1}=q^{2m-2}$ and         $R_{\text{WCLC}}=\big(\frac{q'-z'}{\lfloor\frac{q'-1}{q'-z'}\rfloor}\big)^t=q^t(q-1)^t$          respectively.
Since $(\frac{m}{t})^t<{m\choose t}<(\frac{em}{t})^t$ and $K={m\choose t}q^{2t}$, we have $\frac{t}{eq^2}K^{\frac{1}{t}}<m<\frac{t}{q^2}K^{\frac{1}{t}}$, i.e., $m=\Theta(K^{\frac{1}{t}})$ for fixed $q$ and $t$.
Consequently, we have
$$\frac{F_{\text{C}}}{F_{\text{WCLC}}}=\frac{q^2}{\left(\frac{q}{2}\right)^{\Theta(K^{\frac{1}{t}})}}, \ \ \ \
\frac{R_{\text{C}}}{R_{\text{WCLC}}}=\frac{1}{2^t}.$$
The load of Scheme C is only $\frac{1}{2^t}$ that of the scheme in \cite{WCLC}, and simultaneously the subpacketization of Scheme C has an exponential (if $t=1$) or sub-exponential (if $t>1$) decrease compared to the scheme in \cite{WCLC} when $q>2$ and $m$ is large.

{\it 2) Numerical Comparisons}

Since the schemes in \cite{YTCC,CKSM} are applicable to specific parameters, it is even difficult to find a common number of users between them and the new schemes. So in the following we will give numerical comparisons under approximately the same number of users and user memory ratio.

\begin{figure}
\centering
\includegraphics[width=3.2in]{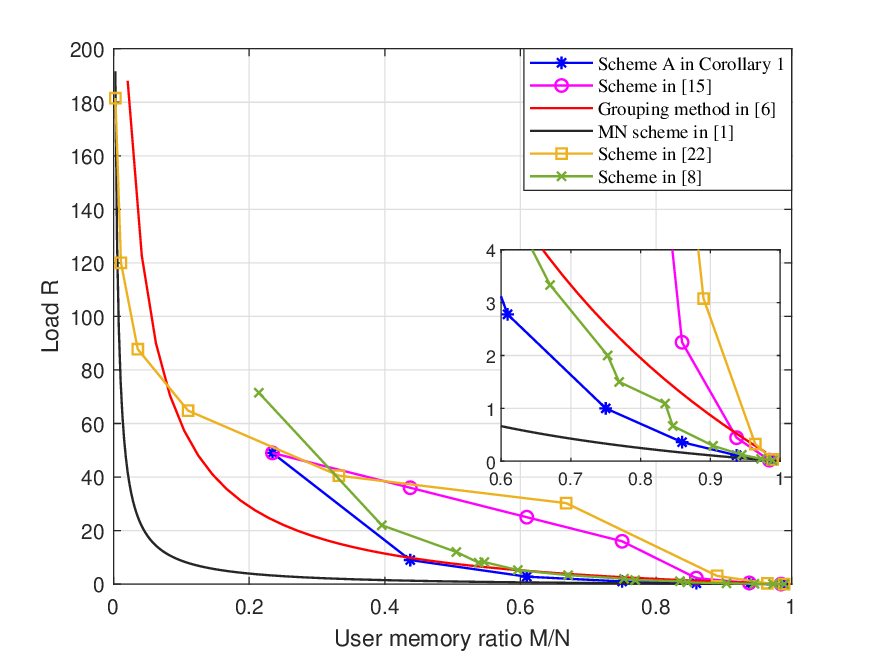}
\includegraphics[width=3.2in]{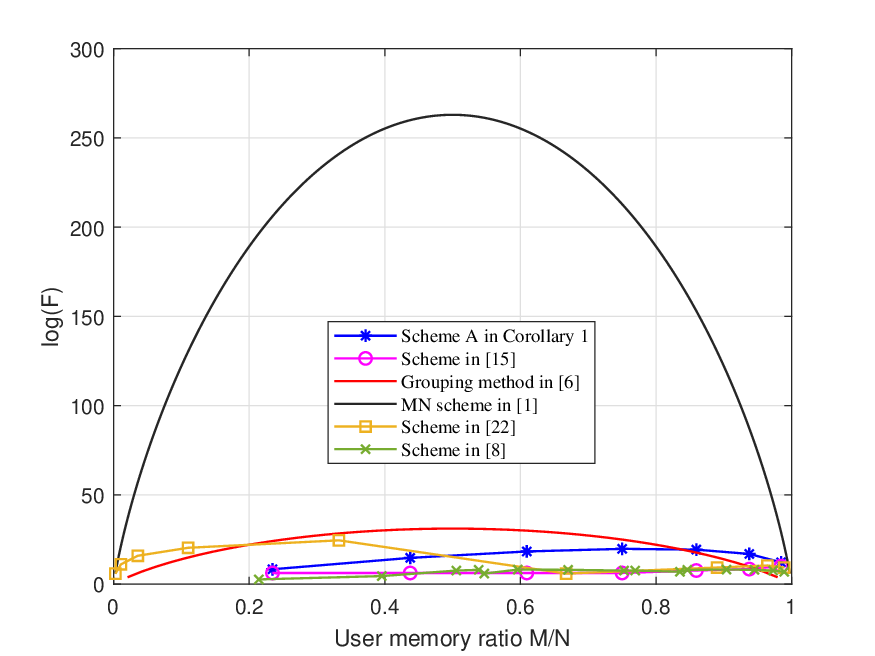}
\caption{The memory-load and memory-subpacketization tradeoffs of Scheme A in Corollary \ref{MNbs} and the schemes in \cite{MN,WCLC,SJTLD,YTCC,CKSM}.}\label{compareRF}
\vspace{-0.8cm}
\end{figure}

The memory-load and memory-subpacketization tradeoffs of Scheme A in Corollary \ref{MNbs} (when $m=4,t=2,q=8$ and $K=384$), the scheme in \cite{WCLC} (when $m=4,t=2,q=8$ and $K=384$), the MN scheme in \cite{MN} (when $K=384$), the grouping method in \cite{SJTLD} (when $K=384, k=48$), the scheme in \cite{CKSM} (when $p=3,k=6,t=1$ and $K=364$) and the scheme in \cite{YTCC} (when $H=14,a=11$ and $K=364$) are shown in Fig. \ref{compareRF}. It can be seen that Scheme A achieves a significantly lower subpacketization than the MN scheme in \cite{MN} and achieves a lower load than the schemes in \cite{WCLC,YTCC}. Moreover, Scheme A achieves a lower load and simultaneously a lower subpacketization than the schemes in \cite{SJTLD,CKSM} for some memory ratios.

The numerical comparisons of Scheme B in Corollary \ref{MNtranbs} and Scheme C in Corollary \ref{thbsg2} with the schemes in \cite{WCLC,SJTLD,CJWY,YTCC,CKSM} are provided in Table \ref{tablecom}. It can be seen that compared to the scheme in \cite{WCLC}, Scheme B, C achieve lower load with higher subpacketization when $m$ is small, and achieve lower load with lower subpacketization when $m$ is large. Compared to the schemes in \cite{SJTLD,CJWY,CKSM}, Scheme B, C can achieve lower load with simultaneously lower subpacketization for some memory ratios.
\begin{table}
       \centering
       \caption{Numerical comparisons  of Scheme B in Corollary \ref{MNtranbs} and Scheme C in Corollary \ref{thbsg2} with the schemes in \cite{WCLC,SJTLD,CJWY,YTCC,CKSM}  }\label{tablecom}
       \small{
       \begin{tabular}{cccccc}
       \toprule
       $K$ &  $M/N$ &Schemes & Parameters & Subpacketization $F$& Load $R$\\
       \midrule
       \multirow{3}*{$216$} & \multirow{3}*{$0.75$} & Scheme B in Corollary \ref{MNtranbs} &  $(m,t,q,z)=(4,2,4,2)$ &$1024$& $2.25$ \\ \cdashline{3-6}

       &  &Scheme in \cite{WCLC} &$(m,t,q,z)=(4,2,6,3)$   & $216$ &$9$\\
       \cdashline{3-6}

       &  &Grouping method in \cite{SJTLD,CJWY}& $(k,t)=(28,21)$ & $8.2883\times 10^6$ & $2.4545$\\ \hdashline

       $220$ & $0.75$ &Scheme in \cite{YTCC}& $(H,a,b,r)=(12,9,1,0)$&$12$&$5.5$\\ \hdashline

       $255$ & $0.7529$ & Scheme in \cite{CKSM}& $(p,k,t,m)=(2,8,1,5)$ & $10795$ & $9$\\
       \hline








       \multirow{3}*{$252$} & \multirow{3}*{$0.2222$} &Scheme B in Corollary \ref{MNtranbs} & $(m,t,q,z)=(7,1,9,2)$ &$9.5660\times 10^6$& $14$ \\ \cdashline{3-6}

       &  & Scheme in \cite{WCLC} &$(m,t,q,z)=(7,1,36,8)$   & $2.1768\times 10^9$ &$28$\\ \cdashline{3-6}

       &  &Grouping method in \cite{SJTLD,CJWY}& $(k,t)=(54,12)$& $1.0290\times 10^{12}$ & $15.0769$\\
       \hdashline

       $286$ & $0.2308$ &Scheme in \cite{YTCC}& $(H,a,b,r)=(13,3,1,0)$&$13$&$55$\\ \hdashline

       $255$ & $0.2471$ & Scheme in \cite{CKSM}& $(p,k,t,m)=(2,8,1,6)$ & $3.0224\times 10^{11}$ & $27.4286$\\
       \hline








       \multirow{3}*{$486$} & \multirow{3}*{$0.5556$} &Scheme C in Corollary \ref{thbsg2}& $(m,t,q)=(4,2,3)$ &$1296$& $9$ \\ \cdashline{3-6}

       &  &Scheme in \cite{WCLC} &$(m,t,q,z)=(4,2,9,3)$   & $729$ &$36$\\ \cdashline{3-6}

       &  &Grouping method in \cite{SJTLD,CJWY}& $(k,t)=(36,20)$& $1.4616\times 10^{10}$ & $10.2857$\\
       \hdashline

       $462$& $0.5671$ & Scheme in \cite{YTCC}&$(H,a,b,r)=(11,5,5,2)$ & $462$ & $10$\\
       \hdashline

       $465$ & $0.5871$ & Scheme in \cite{CKSM}& $(p,k,t,m)=(2,5,2,3)$&$4340$&$19.2$\\
       \hline




       \multirow{3}*{$125$} & \multirow{3}*{$0.2$} &Scheme C in Corollary \ref{thbsg2}&  $(m,t,q)=(5,1,5)$ &$10^5$& $10$ \\ \cdashline{3-6}

       &  &Scheme in \cite{WCLC} & $(m,t,q,z)=(5,1,25,5)$   & $3.9063\times 10^5$ &$20$\\ \cdashline{3-6}

       &  &Grouping method in \cite{SJTLD,CJWY}& $(k,t)=(45,9)$& $7.9755\times 10^{9}$ & $10$\\
       \hdashline

       $120$ & $0.2417$ & Scheme in \cite{YTCC}& $(H,a,b,r)=(16,2,2,0)$&$120$&$15.1667$\\
       \hdashline

       $127$ & $0.2441$ & Scheme in \cite{CKSM}& $(p,k,t,m)=(2,7,1,5)$&$2.2224\times 10^8$&$16$\\




       \bottomrule
  \end{tabular}}
  \vspace{-0.8cm}
\end{table}

\section{conclusion}
\label{conclusion}
In this paper, we first propose a PDA construction by dividing the cache configurations from the $m$-fold Cartesian product of a base PDA into $m$ groups and taking the union of $t$ cache configurations from $t$ different groups, which leads to a coded caching scheme with subpacketization increasing sub-exponentially with the number of users while keeping the load constant for fixed memory ratio. By applying this construction to three existing base PDAs, three new coded caching schemes are obtained, which cover the schemes in \cite{SZG,WCWC} as special cases and can achieve lower load with simultaneously lower subpacketization for some memory ratios compared to existing coded caching schemes. 

\appendices

\section{Proof of Theorem~\ref{thmulpmt}}
\label{prPDAPmt}
\begin{proof}
We will prove that the array $\mathbf{P}_{m,t}$ generated by Construction \ref{constr1} is an ${m\choose t}g_1^t$-$\big({m\choose t}K_1^t, \big.$ $\big.\lambda^t (\frac{F_1}{\lambda})^m,\lambda^t (\frac{F_1}{\lambda})^m(1-(\frac{F_1-Z_1}{F_1})^t),(\frac{F_1}{\lambda})^{m-t}S_1^t\big)$ PDA.

1) Since $\mathbf{P}$ is a base PDA with parameter $\lambda$ and there are $Z_1$ stars in each column of $\mathbf{P}$, there are $(\frac{F_1}{\lambda}-\frac{Z_1}{\lambda})^t(\frac{F_1}{\lambda})^{m-t}\lambda^t$ non-star entries in each column of $\mathbf{P}_{m,t}$ from \eqref{constrPm}, then there are $Z=\lambda^t(\frac{F_1}{\lambda})^m-(\frac{F_1}{\lambda}-\frac{Z_1}{\lambda})^t(\frac{F_1}{\lambda})^{m-t}\lambda^t=\lambda^t (\frac{F_1}{\lambda})^m(1-(\frac{F_1-Z_1}{F_1})^t)$ stars in each column of $\mathbf{P}_{m,t}$. Condition C$1$ of Definition \ref{def-PDA} holds.

2) For any two distinct entries $\mathbf{P}_{m,t}(({\bf f},{\bm \varepsilon}),(\mathcal{T},{\bf b}))$ and $\mathbf{P}_{m,t}(({\bf f}',{\bm \varepsilon}'),(\mathcal{T}',{\bf b}'))$, if $\mathbf{P}_{m,t}(({\bf f},{\bm \varepsilon}),$ $(\mathcal{T},{\bf b}))=\mathbf{P}_{m,t}(({\bf f}',{\bm \varepsilon}'),(\mathcal{T}',{\bf b}'))={\bf e}\neq *$, we will prove that $\mathbf{P}_{m,t}(({\bf f},{\bm \varepsilon}),(\mathcal{T}',{\bf b}'))=\mathbf{P}_{m,t}(({\bf f}', {\bm \varepsilon}'),$  $(\mathcal{T},{\bf b}))=*$.
    Assume that ${\bf f}=(f_1,\ldots,f_m)$, ${\bm \varepsilon}=(\varepsilon_1,\ldots,\varepsilon_t)$,  $\mathcal{T}=\{\delta_1,\ldots,\delta_t\}$ with $\delta_1<\ldots<\delta_t$, ${\bf b}=(b_1,\ldots,b_t)$, $\mathcal{T}\cap [1:m-t+1]=\{\delta_1,\ldots,\delta_{w}\}$, $[m-t+2:m]\setminus \mathcal{T}=\{\sigma_1,\ldots,\sigma_{w-1}\}$ with $\sigma_1<\ldots<\sigma_{w-1}$; ${\bf f}'=(f'_1,\ldots,f'_m)$, ${\bm \varepsilon}'=(\varepsilon'_1,\ldots,\varepsilon'_t)$, $\mathcal{T}'=\{\delta'_1,\ldots,\delta'_t\}$ with $\delta'_1<\ldots<\delta'_t$, ${\bf b}'=(b'_1,\ldots,b'_t)$, $\mathcal{T}'\cap [1:m-t+1]=\{\delta'_1,\ldots,\delta'_{w'}\}$ and $[m-t+2:m]\setminus \mathcal{T}'=\{\sigma'_1,\ldots,\sigma'_{w'-1}\}$ with $\sigma'_1<\ldots<\sigma'_{w'-1}$.
    From \eqref{constrPm} we have $\mathbf{P}(f_{\delta_h},b_h)\neq*$ and $\mathbf{P}(f'_{\delta'_h},b'_h)\neq*$ for any $h\in[1:t]$. Since $\mathbf{P}$ is a base PDA with parameter $\lambda$, we have $\mathbf{P}(f_{\delta_h}+(\varepsilon_h-1)\frac{F_1}{\lambda},b_h)\neq *$ and $\mathbf{P}(f'_{\delta'_h}+(\varepsilon'_h-1)\frac{F_1}{\lambda},b'_h)\neq *$ from Condition C4 of Definition \ref{def-basePDA}. Since $\mathcal{B}_1, \mathcal{B}_2,\ldots,\mathcal{B}_{\frac{F_1}{\lambda}}$ is a uniform partition of $[1:S_1]$, there exist unique $l_h,\mu_h,l'_h,\mu'_h$ such that
        \begin{alignat}{2}
        \mathbf{P}(f_{\delta_h}+(\varepsilon_h-1)\frac{F_1}{\lambda},b_h)&=&\mathcal{B}_{l_h}[\mu_h], \label{lu} \\ \mathbf{P}(f'_{\delta'_h}+(\varepsilon'_h-1)\frac{F_1}{\lambda},b'_h)&=&\mathcal{B}_{l'_h}[\mu'_h],\label{lu1}
        \end{alignat}
        for any $h\in[1:t]$.
    
    If $\mathcal{T}=\mathcal{T}'$, which implies $\delta_h=\delta'_h$ for any $h\in[1:t]$, $w=w'$ and $\sigma_h=\sigma'_h$ for any $h\in[1:w-1]$. From \eqref{constre} we have
    \begin{equation}
    \label{e3}
    e_{\delta_h}=\mathcal{B}_{l_h}[\mu_1]=\mathcal{B}_{l'_h}[\mu'_1] \ \ \text{for} \ \ h\in[1:w],
    \end{equation}
    \begin{equation}
    \label{e4}
    e_{\delta_h}=\mathcal{B}_{l_h}[\mu_h]=\mathcal{B}_{l'_h}[\mu'_h] \ \ \text{for} \ \ h\in[w+1:t],
    \end{equation}
    \begin{equation}
    \label{e2}
    e_{\sigma_h}=\mathcal{B}_{f_{\sigma_h}}[\mu_{h+1}]=\mathcal{B}_{f'_{\sigma_h}}[\mu'_{h+1}] \ \ \text{for} \ \ h\in[1:w-1],
    \end{equation}
    \begin{equation}
    \label{e1}
    e_i=\mathcal{B}_{f_i}[\mu_1]=\mathcal{B}_{f'_i}[\mu'_1]\ \ \text{for} \ \ i\in[1:m-t+1]\setminus \mathcal{T}.
    \end{equation}
    From \eqref{e3}, \eqref{e4} and \eqref{e2} we have $\mu_h=\mu'_h, l_h=l'_h$ for any $h\in[1:t]$, leading to  $\mathbf{P}(f_{\delta_h}+(\varepsilon_h-1)\frac{F_1}{\lambda},b_h)=\mathbf{P}(f'_{\delta_h}+(\varepsilon'_h-1)\frac{F_1}{\lambda},b'_h)$ from \eqref{lu} and \eqref{lu1}, which implies $f_{\delta_h}+(\varepsilon_h-1)\frac{F_1}{\lambda}=f'_{\delta_h}+(\varepsilon'_h-1)\frac{F_1}{\lambda},b_h=b'_h$ or $\mathbf{P}(f_{\delta_h}+(\varepsilon_h-1)\frac{F_1}{\lambda},b'_h)=\mathbf{P}(f'_{\delta_h}+(\varepsilon'_h-1)\frac{F_1}{\lambda},b_h)=*$ from Condition C3 of Definition \ref{def-PDA}.
    If $f_{\delta_h}+(\varepsilon_h-1)\frac{F_1}{\lambda}=f'_{\delta_h}+(\varepsilon'_h-1)\frac{F_1}{\lambda},b_h=b'_h$ hold for any $h\in[1:t]$, since $f_{\delta_h}, f'_{\delta_h} \in[1:\frac{F_1}{\lambda}]$, we have $f_{\delta_h}=f'_{\delta_h}, \varepsilon_h=\varepsilon'_h$. Since $f_i=f'_i$ for any $i\in[1:m]\setminus \mathcal{T}$ from \eqref{e2} and \eqref{e1}, we have ${\bf f}={\bf f}', {\bm \varepsilon}={\bm \varepsilon}'$ and ${\bf b}={\bf b}'$, which contradicts the hypothesis that $\mathbf{P}_{m,t}(({\bf f},{\bm \varepsilon}),(\mathcal{T},{\bf b}))$ and $\mathbf{P}_{m,t}(({\bf f}',{\bm \varepsilon}'),(\mathcal{T}',{\bf b}'))$ are two distinct entries. Hence, there exists $h_0\in[1:t]$, such that $\mathbf{P}(f_{\delta_{h_0}}+(\varepsilon_{h_0}-1)\frac{F_1}{\lambda},b'_{h_0})=\mathbf{P}(f'_{\delta_{h_0}}+(\varepsilon'_{h_0}-1)\frac{F_1}{\lambda},b_{h_0})=*$, leading to $\mathbf{P}(f_{\delta_{h_0}},b'_{h_0})=\mathbf{P}(f'_{\delta_{h_0}},b_{h_0})=*$ from Condition C4 of Definition \ref{def-basePDA}, which implies $\mathbf{P}_{m,t}(({\bf f},{\bm \varepsilon}),(\mathcal{T}',{\bf b}'))=\mathbf{P}_{m,t}(({\bf f}', {\bm \varepsilon}'),$  $(\mathcal{T},{\bf b}))=*$ from \eqref{constrPm}.

    If $\mathcal{T}\neq \mathcal{T}'$, then there exists $\delta_{\alpha}\in\mathcal{T}$ such that $\delta_{\alpha}\notin\mathcal{T}'$.
        If $\alpha\in[1:m-t+1]$, we have
        $e_{\delta_{\alpha}}=\mathcal{B}_{l_{\alpha}}[\mu_1]=\mathcal{B}_{f'_{\delta_{\alpha}}}[\mu'_1]$
        from \eqref{constre}, which implies $l_{\alpha}=f'_{\delta_{\alpha}}$.
        If $\alpha\in[m-t+2:m]$, we have $\delta_{\alpha}\in[m-t+2:m]\setminus \mathcal{T}'=\{\sigma'_1,\ldots,\sigma'_{w'-1}\}$, assume that $\delta_{\alpha}=\sigma'_{\beta}$, then
            $e_{\delta_{\alpha}}=\mathcal{B}_{l_\alpha}[\mu_\alpha]=\mathcal{B}_{f'_{\delta_{\alpha}}}[\mu'_{\beta+1}]$
            from \eqref{constre}, which also implies $l_\alpha=f'_{\delta_{\alpha}}$.
        Consequently, we have
            $\mathbf{P}(f_{\delta_{\alpha}}+(\varepsilon_\alpha-1)\frac{F_1}{\lambda},b_{\alpha})=\mathcal{B}_{f'_{\delta_{\alpha}}}[\mu_{\alpha}]$ from \eqref{lu}, which implies that the $f'_{\delta_{\alpha}}$-th row is a star row for the non-star entry $\mathbf{P}(f_{\delta_{\alpha}}+(\varepsilon_\alpha-1)\frac{F_1}{\lambda},b_{\alpha})$, i.e., $\mathbf{P}(f'_{\delta_{\alpha}},b_{\alpha})=*$. Hence, we have $\mathbf{P}_{m,t}(({\bf f}', {\bm \varepsilon}'),$  $(\mathcal{T},{\bf b}))=*$ from \eqref{constrPm}.
        Similarly, we can prove that $\mathbf{P}_{m,t}(({\bf f},{\bm \varepsilon}),(\mathcal{T}',{\bf b}'))=*$.

    Condition C$3$ of Definition \ref{def-PDA} holds.

3) For any non-star entry in $\mathbf{P}_{m,t}$, say $\mathbf{P}_{m,t}(({\bf f},{\bm \varepsilon}),(\mathcal{T},{\bf b}))={\bf e}$, we will prove that ${\bf e}$ appears exactly ${m\choose t}g_1^t$ times in $\mathbf{P}_{m,t}$ by proving the following two statements.
\begin{itemize}
\item[a)]  For any $\mathcal{T}'\in{[1:m]\choose t}$ with $\mathcal{T}'\neq \mathcal{T}$, there exist ${\bf f}'\in [1:\frac{F_1}{\lambda}]^m$, ${\bm \varepsilon}'\in[1:\lambda]^t$ and ${\bf b}'\in[1:K_1]^t$ such that $\mathbf{P}_{m,t}(({\bf f}',{\bm \varepsilon}'),(\mathcal{T}',{\bf b}'))={\bf e}$;
\item[b)]  ${\bf e}$ occurs exactly $g_1^t$ times in the columns indexed by $(\mathcal{T},{\bf c})$ where ${\bf c}\in[1:K_1]^t$.
\end{itemize}

    Assume that ${\bf f}=(f_1,\ldots,f_m)$, ${\bm \varepsilon}=(\varepsilon_1,\ldots, \varepsilon_t)$, $\mathcal{T}=\{\delta_1,\ldots,\delta_t\}$ with $\delta_1<\ldots<\delta_t$, ${\bf b}=(b_1,b_2,\ldots,b_t)$, $\mathcal{T}\cap [1:m-t+1]=\{\delta_1,\ldots,\delta_{w}\}$ and $[m-t+2:m]\setminus \mathcal{T}=\{\sigma_1,\ldots,\sigma_{w-1}\}$ with $\sigma_1<\ldots<\sigma_{w-1}$. For any $h\in[1:t]$, we have $\mathbf{P}(f_{\delta_h},b_h)\neq *$ from \eqref{constrPm}, then $\mathbf{P}(f_{\delta_h}+(\varepsilon_h-1)\frac{F_1}{\lambda},b_h)\neq *$ from Condition C4 of Definition \ref{def-basePDA}. Since $\mathcal{B}_1,\ldots,\mathcal{B}_{\frac{F_1}{\lambda}}$ is a uniform partition of $[1:S_1]$, there exist unique $l_h$ and $\mu_h$ such that
    \begin{equation}
    \label{fblm}
    \mathbf{P}(f_{\delta_h}+(\varepsilon_h-1)\frac{F_1}{\lambda},b_h)=\mathcal{B}_{l_h}[\mu_h] \ \ \text{for any} \ \ h\in[1:t].
    \end{equation}
    Let us consider the first statement. If $|\mathcal{T}'\cap\mathcal{T}|=t-1$, assume that $\mathcal{T}'=\{\delta'_1,\ldots,\delta'_t\}$ with $\delta'_1<\ldots<\delta'_t$,  $\mathcal{T}\setminus\mathcal{T}'=\{\delta_{\alpha}\}$ and $\mathcal{T}'\setminus\mathcal{T}=\{\delta'_\beta\}$. Let
    \begin{equation}
    \label{falfa}
    f'_{\delta_{\alpha}}=l_{\alpha}.
    \end{equation}
    
    {\it Case 1:} If $\delta_{\alpha},\delta'_\beta\in[1:m-t+1]$, then $[m-t+2:m]\setminus\mathcal{T}'=[m-t+2:m]\setminus\mathcal{T}=\{\sigma_1,\ldots,\sigma_{w-1}\}$.
       If $\delta_{\alpha}>\delta'_\beta$, then we have
           \begin{equation}
           \label{newT}
           \delta'_h=\begin{cases}
           \delta_h,\ \ & \text{if} \ \ h\in[1:\beta-1]\cup[\alpha+1:t],\\
           \delta'_\beta, \ \ & \text{if} \ \ h=\beta,\\
           \delta_{h-1}, \ \ &\text{if} \ \ h\in[\beta+1:\alpha].
           \end{cases}
           \end{equation}
           Let
           \begin{alignat}{3}
           &f'_i&=&f_i \ \ \text{for any} \ \ i\in[1:m]\setminus\{\delta'_\beta,\delta'_{\beta+1},\ldots,\delta'_{\alpha},\delta_{\alpha}\}, \label{samef}\\
           &\varepsilon'_h&=&\varepsilon_h \ \ \text{for any} \ \  h\in[1:\beta-1]\cup[\alpha+1:t],\label{sameep}\\
           &b'_h&=&b_h \ \ \text{for any} \ \  h\in[1:\beta-1]\cup[\alpha+1:t],\label{sameb}
           \end{alignat}
           then for any $h\in[1:\beta-1]\cup[\alpha+1:t]$, we have
           \begin{equation}
           \label{samefeb}
           \mathbf{P}(f'_{\delta_h}+(\varepsilon'_h-1)\frac{F_1}{\lambda},b'_h)=\mathbf{P}(f_{\delta_h}+(\varepsilon_h-1)\frac{F_1}{\lambda},b_h)=\mathcal{B}_{l_h}[\mu_h].
           \end{equation}
           Assume that $\mathcal{B}_{f_{\delta'_\beta}}[\mu_{\beta}]=s$, there exist $j\in[1:F_1]$ and $k\in[1:K_1]$ such that $\mathbf{P}(j,k)=s$, let $f'_{\delta'_\beta}=\langle j\rangle_{\frac{F_1}{\lambda}}$, $\varepsilon'_\beta=\frac{j-f'_{\delta'_\beta}}{\frac{F_1}{\lambda}}+1$ and $b'_{\beta}=k$, then
           \begin{equation}
           \label{fbbeta1}
           \mathbf{P}(f'_{\delta'_{\beta}}+(\varepsilon'_\beta-1)\frac{F_1}{\lambda},b'_{\beta})=\mathcal{B}_{f_{\delta'_\beta}}[\mu_{\beta}].
           \end{equation}
           For any $h\in[\beta+1:\alpha]$, assume that $\mathcal{B}_{l_{h-1}}[\mu_{h}]=s_h$, there exist $j_h\in[1:F_1]$ and $k_h\in[1:K_1]$ such that $\mathbf{P}(j_h,k_h)=s_h$, let $f'_{\delta'_{h}}=\langle j_h\rangle_{\frac{F_1}{\lambda}}$, $\varepsilon'_h=\frac{j_h-f'_{\delta'_{h}}}{\frac{F_1}{\lambda}}+1$, and $b'_{h}=k_h$, then
           \begin{equation}
           \label{fbdeltai+}
           \mathbf{P}(f'_{\delta'_{h}}+(\varepsilon'_h-1)\frac{F_1}{\lambda},b'_{h})=\mathcal{B}_{l_{h-1}}[\mu_{h}]  \ \text{for any} \ \ h\in[\beta+1:\alpha].
           \end{equation}
           From \eqref{samef} \eqref{sameep} and \eqref{sameb} we have
           $\mathbf{P}(f'_{\delta'_h},b'_h)=\mathbf{P}(f_{\delta_h},b_h)\neq*$ for any $h\in[1:\beta-1]\cup[\alpha+1:t]$. From \eqref{fbbeta1} and \eqref{fbdeltai+} we have $\mathbf{P}(f'_{\delta'_h}+(\varepsilon'_h-1)\frac{F_1}{\lambda},b'_h)\neq *$ for any $h\in[\beta:\alpha]$, which implies $\mathbf{P}(f'_{\delta'_h},b'_h)\neq *$ from Condition C4 of Definition \ref{def-basePDA}.  Let ${\bf f}'=(f'_1,\ldots,f'_m)$, ${\bm \varepsilon}'=(\varepsilon'_1,\ldots,\varepsilon'_t)$, ${\bf b}'=(b'_1,\ldots,b'_t)$, then we have $\mathbf{P}_{m,t}(({\bf f}',{\bm \varepsilon}'),(\mathcal{T}',{\bf b}'))\neq *$ from \eqref{constrPm}. Assume that $\mathbf{P}(f'_{\delta'_h}+(\varepsilon'_h-1)\frac{F_1}{\lambda},b'_h)=\mathcal{B}_{l'_h}[\mu'_h]$ for any $h\in[1:t]$, then from \eqref{samefeb} \eqref{fbbeta1} and \eqref{fbdeltai+} we have \begin{equation}
           \label{mu1}
           \mu'_h=\mu_h, \ \ \text{for any} \ \ h\in[1:t]
           \end{equation}
           and
           \begin{equation}
           \label{l1}
           l'_h=\begin{cases}
           l_h,\ \ &\text{if} \ \ h\in[1:\beta-1]\cup[\alpha+1:t],\\
           f_{\delta'_\beta},\ \ &\text{if} \ \ h=\beta,\\
           l_{h-1},\ \ &\text{if} \ \ h\in[\beta+1:\alpha].
           \end{cases}
           \end{equation}
           Assume that $\mathbf{P}_{m,t}(({\bf f}',{\bm \varepsilon}'),(\mathcal{T}',{\bf b}'))={\bf e}'=(e'_1,\ldots,e'_m)$, we have
           \begin{alignat*}{6}
            &e'_{\delta_h}\overset{\eqref{newT}}{=}e'_{\delta'_h}\overset{ \eqref{constre}}{=}\mathcal{B}_{l'_h}[\mu'_1]\overset{  \eqref{mu1},\eqref{l1}}{=}\mathcal{B}_{l_h}[\mu_1]\overset{ \eqref{constre}}{=}e_{\delta_h}, \ \ h\in[1:\beta-1]\cup[\alpha+1:w],\\
            &e'_{\delta_{h}}\overset{\eqref{newT}}{=}e'_{\delta'_{h+1}}\overset{  \eqref{constre}}{=}\mathcal{B}_{l'_{h+1}}[\mu'_1]\overset{ \eqref{mu1},\eqref{l1}}{=}\mathcal{B}_{l_h}[\mu_1]\overset{  \eqref{constre}}{=}e_{\delta_{h}}, \ \ h\in[\beta:\alpha-1],\\
            &e'_{\delta_{h}}\overset{\eqref{newT}}{=}e'_{\delta'_{h}}\overset{  \eqref{constre}}{=}\mathcal{B}_{l'_{h}}[\mu'_h]\overset{ \eqref{mu1},\eqref{l1}}{=}\mathcal{B}_{l_h}[\mu_h]\overset{  \eqref{constre}}{=}e_{\delta_{h}}, \ \ h\in[w+1:t],\\
            &e'_{\delta'_\beta}\overset{\eqref{constre}}{=}\mathcal{B}_{l'_\beta}[\mu'_1]
            \overset{\eqref{mu1},\eqref{l1}}{=}\mathcal{B}_{f_{\delta'_\beta}}[\mu_1]\overset{  \eqref{constre}}{=}e_{\delta'_\beta},\\
            &e'_{\delta_\alpha}\overset{\eqref{constre}}{=}\mathcal{B}_{f'_{\delta_\alpha}}[\mu'_1]
            \overset{\eqref{falfa},\eqref{mu1}}{=}\mathcal{B}_{l_\alpha}[\mu_1]\overset{\eqref{constre}}{=}e_{\delta_\alpha},\\
            &e'_i\overset{\eqref{constre}}{=}\mathcal{B}_{f'_i}[\mu'_1]\overset{\eqref{samef},\eqref{mu1}}{=}\mathcal{B}_{f_i}[\mu_1]\overset{\eqref{constre}}{=}e_i, \ \ i\in[1:m-t+1]\setminus(\mathcal{T}\cup\mathcal{T}'),\\
            &e'_{\sigma_h}\overset{\eqref{constre}}{=}\mathcal{B}_{f'_{\sigma_h}}[\mu'_{h+1}]\overset{\eqref{samef},\eqref{mu1}}{=}\mathcal{B}_{f_{\sigma_h}}[\mu_{h+1}]\overset{\eqref{constre}}{=}e_{\sigma_h}, \ \  h\in[1:w-1],
            \end{alignat*}
           i.e., ${\bf e}'={\bf e}$. Therefore, $\mathbf{P}_{m,t}(({\bf f}',{\bm \varepsilon}'),(\mathcal{T}',{\bf b}'))={\bf e}$.
           If $\delta_{\alpha}<\delta'_{\beta}$, it can be similarly proved that there exist ${\bf f}'\in[1:\frac{F_1}{\lambda}]^m$, ${\bm \varepsilon}'\in[1:\lambda]^t$ and ${\bf b}'\in[1:K_1]^t$ such that $\mathbf{P}_{m,t}(({\bf f}',{\bm \varepsilon}'),(\mathcal{T}',{\bf b}'))={\bf e}$.

    {\it Case 2:} If $\delta_{\alpha},\delta'_\beta\in[m-t+2:m]$, then $|[m-t+2:m]\setminus \mathcal{T}'|=|[m-t+2:m]\setminus \mathcal{T}|=w-1$. Since $\delta'_\beta\in[m-t+2:m]\setminus \mathcal{T}=\{\sigma_1,\ldots,\sigma_{w-1}\}$ and $\delta_{\alpha}\in\mathcal{T}\cap[m-t+2:m]$, assume that
        \begin{equation}
        \label{sigmad}
        \delta'_\beta=\sigma_{\gamma_1},\ \sigma_{\gamma_2}<\delta_{\alpha}<\sigma_{\gamma_2+1}.
        \end{equation}
        If $\delta_{\alpha}>\delta'_\beta$, we have \eqref{newT} and $\sigma_{\gamma_1}<\sigma_{\gamma_2+1}$, then $\sigma_{\gamma_1}\leq \sigma_{\gamma_2}$, which implies $\gamma_1\leq\gamma_2\leq w-1$. Assume that $[m-t+2:m]\setminus\mathcal{T}'=\{\sigma'_1,\ldots,\sigma'_{w-1}\}$ with $\sigma'_1<\ldots<\sigma'_{w-1}$, we have
            \begin{equation}
            \label{sigma1}
            \sigma'_h=\begin{cases}
            \sigma_h,\ \,&\text{if} \ \ h\in[1:\gamma_1-1]\cup[\gamma_2+1:w-1],\\
            \sigma_{h+1}, \ \ &\text{if} \ \ h\in[\gamma_1:\gamma_2-1],\\
            \delta_{\alpha},\ \ &\text{if} \ \ h=\gamma_2.
            \end{cases}
            \end{equation}
            Let
            \begin{alignat}{3}
            &f'_i=f_i \ \ \text{for any} \ \ i\in[1:m]\setminus \{\delta_{\gamma_1+1},\ldots,\delta_{\gamma_2+1},\delta'_\beta,\delta_{\alpha}\},\label{samef2}\\
            &b'_h=b_h, \varepsilon'_h=\varepsilon_h \ \ \text{for any} \ \ h\in[1:\gamma_1]\cup[\gamma_2+2:\beta-1]\cup[\alpha+1:t],\label{sameb2}\\
            &b'_h=b_{h-1}, \varepsilon'_h=\varepsilon_{h-1} \ \ \text{for any} \ \ h\in[\beta+1:\alpha],\label{difb2}
            \end{alignat}
            then for any $h\in[1:\gamma_1]\cup[\gamma_2+2:\beta-1]\cup[\alpha+1:t]$ we have
            \begin{equation}
            \label{sameef1}
            \mathbf{P}(f'_{\delta'_h}+(\varepsilon'_h-1)\frac{F_1}{\lambda},b'_h)\overset{\eqref{newT},\eqref{samef2},\eqref{sameb2}}{=}\mathbf{P}(f_{\delta_h}+(\varepsilon_h-1)\frac{F_1}{\lambda},b_h)=\mathcal{B}_{l_h}[\mu_h]; \end{equation}
            for any $h\in[\beta+1:\alpha]$ we have
            \begin{equation}
            \label{sameef2}
            \mathbf{P}(f'_{\delta'_h}+(\varepsilon'_h-1)\frac{F_1}{\lambda},b'_h)\overset{\eqref{newT},\eqref{samef2},\eqref{difb2}}{=}\mathbf{P}(f_{\delta_{h-1}}+(\varepsilon_{h-1}-1)\frac{F_1}{\lambda},b_{h-1})=\mathcal{B}_{l_{h-1}}[\mu_{h-1}]. 
        \end{equation}
            For any $h\in[\gamma_1+1:\gamma_2]$, assume that $\mathcal{B}_{l_h}[\mu_{h+1}]=s_h$, there exist $j_h\in[1:F_1]$ and $k_h\in[1:K_1]$ such that $\mathbf{P}(j_h,k_h)=s_h$. Let $f'_{\delta'_{h}}=\langle j_h\rangle_{\frac{F_1}{\lambda}}$, $\varepsilon'_h=\frac{j_h-f'_{\delta'_{h}}}{\frac{F_1}{\lambda}}+1$ and $b'_{h}=k_h$, then
            \begin{equation}
            \label{fbdeltaip}
            \mathbf{P}(f'_{\delta'_{h}}+(\varepsilon'_h-1)\frac{F_1}{\lambda},b'_{h})=\mathcal{B}_{l_h}[\mu_{h+1}] \ \text{for any} \ \ h\in[\gamma_1+1:\gamma_2].
            \end{equation}
            Assume that $\mathcal{B}_{l_{\gamma_2+1}}[\mu_{\alpha}]=s$, there exist $j\in[1:F_1]$ and $k\in[1:K_1]$ such that $\mathbf{P}(j,k)=s$. Let $f'_{\delta'_{\gamma_2+1}}=\langle j\rangle_{\frac{F_1}{\lambda}}$, $\varepsilon'_{\gamma_2+1}=\frac{j-f'_{\delta'_{\gamma_2+1}}}{\frac{F_1}{\lambda}}+1$ and $b'_{\gamma_2+1}=k$, then
            \begin{equation}
            \label{fdeltabetap}
            \mathbf{P}(f'_{\delta'_{\gamma_2+1}}+(\varepsilon'_{\gamma_2+1}-1)\frac{F_1}{\lambda},b'_{\gamma_2+1})=\mathcal{B}_{l_{\gamma_2+1}}[\mu_{\alpha}].
            \end{equation}
            Assume that $\mathcal{B}_{f_{\delta'_\beta}}[\mu_{\gamma_1+1}]=s'$, there exist $j'\in[1:F_1]$ and $k'\in[1:K_1]$ such that $\mathbf{P}(j',k')=s'$. Let $f'_{\delta'_\beta}=\langle j'\rangle_{\frac{F_1}{\lambda}}$, $\varepsilon'_\beta=\frac{j'-f'_{\delta'_\beta}}{\frac{F_1}{\lambda}}+1$ and $b'_{\beta}=k'$, then
            \begin{equation}
            \label{fbgamap}
            \mathbf{P}(f'_{\delta'_{\beta}}+(\varepsilon'_\beta-1)\frac{F_1}{\lambda},b'_{\beta})=\mathcal{B}_{f_{\delta'_\beta}}[\mu_{\gamma_1+1}].
            \end{equation}
        From \eqref{sameef1}, \eqref{sameef2}, \eqref{fbdeltaip}, \eqref{fdeltabetap} and \eqref{fbgamap} we have $\mathbf{P}(f'_{\delta'_h}+(\varepsilon'_h-1)\frac{F_1}{\lambda},b'_h)\neq *$ for any $h\in[1:t]$, which implies $\mathbf{P}(f'_{\delta'_h},b'_h)\neq *$ from Condition C4 of Definition \ref{def-basePDA}.
        Let ${\bf f}'=(f'_1,\ldots,f'_m)$, ${\bm \varepsilon}'=(\varepsilon'_1,\ldots,\varepsilon'_t)$, ${\bf b}'=(b'_1,\ldots,b'_t)$, then we have $\mathbf{P}_{m,t}(({\bf f}',{\bm \varepsilon}'),(\mathcal{T}',{\bf b}'))\neq *$ from \eqref{constrPm}. Assume that $\mathbf{P}(f'_{\delta'_h}+(\varepsilon'_h-1)\frac{F_1}{\lambda},b'_h)=\mathcal{B}_{l'_h}[\mu'_h]$ for any $h\in[1:t]$, then from \eqref{sameef1}, \eqref{sameef2}, \eqref{fbdeltaip}, \eqref{fdeltabetap} and \eqref{fbgamap} we have
           \begin{equation}
           \label{mu3}
           \mu'_h=\begin{cases}
           \mu_h, \ \ &\text{if} \ \ h\in[1:\gamma_1]\cup[\gamma_2+2:\beta-1]\cup[\alpha+1:t],\\
           \mu_{h+1}, \ \ &\text{if} \ \ h\in[\gamma_1+1:\gamma_2],\\
           \mu_{h-1}, \ \ &\text{if} \ \ h\in[\beta+1:\alpha],\\
           \mu_{\alpha},\ \ &\text{if} \ \ h=\gamma_2+1,\\
           \mu_{\gamma_1+1},\ \ &\text{if} \ \ h=\beta
           \end{cases}
           \end{equation}
           and
           \begin{equation}
           \label{l3}
           l'_h=\begin{cases}
           l_h,\ \ &\text{if} \ \ h\in[1:\beta-1]\cup[\alpha+1:t],\\
           l_{h-1}, \ \ &\text{if} \ \ h\in[\beta+1:\alpha],\\
           f_{\delta'_\beta},\ \ &\text{if} \ \ h=\beta.
           \end{cases}
           \end{equation}
          Assume that $\mathbf{P}_{m,t}(({\bf f}',{\bm \varepsilon}'),(\mathcal{T}',{\bf b}'))={\bf e}'=(e'_1,\ldots,e'_m)$, we have
           \begin{alignat*}{5}
           &e'_{\delta_h}\overset{\eqref{newT}}{=}e'_{\delta'_h}\overset{\eqref{constre}}{=}\mathcal{B}_{l'_h}[\mu'_1]\overset{ \eqref{mu3},\eqref{l3}}{=}\mathcal{B}_{l_h}[\mu_1]\overset{\eqref{constre}}{=}e_{\delta_h}, h\in[1:w], \\
           &e'_{\delta_{h}}\overset{\eqref{newT}}{=}e'_{\delta'_{h}}\overset{\eqref{constre}}{=}\mathcal{B}_{l'_{h}}[\mu'_h]\overset{ \eqref{mu3},\eqref{l3}}{=}\mathcal{B}_{l_h}[\mu_h]\overset{\eqref{constre}}{=}e_{\delta_{h}}, h\in[w+1:\beta-1]\cup[\alpha+1:t], \\
           &e'_{\delta_{h}}\overset{\eqref{newT}}{=}e'_{\delta'_{h+1}}\overset{\eqref{constre}}{=}\mathcal{B}_{l'_{h+1}}[\mu'_{h+1}]\overset{  \eqref{mu3},\eqref{l3}}{=}\mathcal{B}_{l_h}[\mu_h]\overset{\eqref{constre}}{=}e_{\delta_{h}},h\in [\beta:\alpha-1], \\
           &e'_{\delta_{\alpha}}\overset{\eqref{sigma1}}{=}e'_{\sigma'_{\gamma_2}}\overset{\eqref{constre}}{=}\mathcal{B}_{f'_{\sigma'_{\gamma_2}}}[\mu'_{\gamma_2+1}]\overset{\eqref{sigma1}}{=}\mathcal{B}_{f'_{\delta_\alpha}}[\mu'_{\gamma_2+1}]\overset{  \eqref{falfa},\eqref{mu3}}{=}\mathcal{B}_{l_{\alpha}}[\mu_{\alpha}]\overset{\eqref{constre}}{=}e_{\delta_{\alpha}}, \\
           &e'_{\delta'_\beta}\overset{\eqref{constre}}{=}\mathcal{B}_{l'_{\beta}}[\mu'_{\beta}]\overset{\eqref{mu3},\eqref{l3}}{=}\mathcal{B}_{f_{\delta'_\beta}}[\mu_{\gamma_1+1}]\overset{\eqref{sigmad}}{=}\mathcal{B}_{f_{\sigma_{\gamma_1}}}[\mu_{\gamma_1+1}]\overset{ \eqref{constre}}{=}e_{\sigma_{\gamma_1}}\overset{\eqref{sigmad}}{=}e_{\delta'_\beta},\\
           &e'_i\overset{\eqref{constre}}{=}\mathcal{B}_{f'_i}[\mu'_1]\overset{\eqref{samef2},\eqref{mu3}}{=}\mathcal{B}_{f_i}[\mu_1]\overset{\eqref{constre}}{=}e_i,i\in[1:m-t+1]\setminus(\mathcal{T}\cup\mathcal{T}'),\\
           &e'_{\sigma_h}\overset{\eqref{sigma1}}{=}e'_{\sigma'_h}\overset{\eqref{constre}}{=}\mathcal{B}_{f'_{\sigma'_h}}[\mu'_{h+1}]\overset{  \eqref{samef2},\eqref{mu3}}{=}\mathcal{B}_{f_{\sigma_h}}[\mu_{h+1}]\overset{\eqref{constre}}{=}e_{\sigma_h},h\in[1:\gamma_1-1]\cup[\gamma_2+1:w-1],\\
           &e'_{\sigma_h}\overset{\eqref{sigma1}}{=}e'_{\sigma'_{h-1}}\overset{\eqref{constre}}{=}\mathcal{B}_{f'_{\sigma'_{h-1}}}[\mu'_{h}]\overset{\eqref{sigma1}}{=}\mathcal{B}_{f'_{\sigma_{h}}}[\mu'_{h}]\overset{  \eqref{samef2},\eqref{mu3}}{=}\mathcal{B}_{f_{\sigma_h}}[\mu_{h+1}]\overset{\eqref{constre}}{=}e_{\sigma_h}, h\in[\gamma_1+1:\gamma_2], \\
           \end{alignat*}
          i.e., ${\bf e}'={\bf e}$. Therefore, $\mathbf{P}_{m,t}(({\bf f}',{\bm \varepsilon}'),(\mathcal{T}',{\bf b}'))={\bf e}$.
       If $\delta_{\alpha}<\delta'_{\beta}$, it can be similarly proved that there exist ${\bf f}'\in[1:\frac{F_1}{\lambda}]^m$, ${\bm \varepsilon}'\in[1:\lambda]^t$ and ${\bf b}'\in[1:K_1]^t$ such that $\mathbf{P}_{m,t}(({\bf f}',{\bm \varepsilon}'),(\mathcal{T}',{\bf b}'))={\bf e}$.

    {\it Case 3:} If $\delta_{\alpha}\in[1:m-t+1]$ and $\delta'_\beta\in[m-t+2:m]$, then $|[m-t+2:m]\setminus\mathcal{T}'|=w-2$, assume that $[m-t+2:m]\setminus\mathcal{T}'=\{\sigma'_1,\ldots,\sigma'_{w-2}\}$ with $\sigma'_1<\ldots<\sigma'_{w-2}$.
        Since $\delta'_\beta\in[m-t+2:m]\setminus \mathcal{T}=\{\sigma_1,\ldots,\sigma_{w-1}\}$, assume that
        \begin{equation}
        \label{sigmad1}
        \delta'_\beta=\sigma_{\gamma}.
        \end{equation}
        Then we have
        \begin{equation}
        \label{newT1}
        \delta'_h=\begin{cases}
        \delta_h, \ \ \text{if} \ \ h\in[1:\alpha-1]\cup[\beta+1:t],\\
        \delta_{h+1}, \ \ \text{if} \ \ h\in[\alpha:\beta-1],\\
        \delta'_\beta, \ \ \text{if} \ \  h=\beta.
        \end{cases}
        \end{equation}
         and
            \begin{equation}
            \label{sigma3}
            \sigma'_h=\begin{cases}
            \sigma_h,\ \,&\text{if} \ \ h\in[1:\gamma-1],\\
            \sigma_{h+1}, \ \ &\text{if} \ \ h\in[\gamma:w-2].
            \end{cases}
            \end{equation}
        If $\alpha\leq\gamma+1$, let
            \begin{alignat}{3}
        	&f'_i=f_i \ \ \text{for any} \ \ i\in[1:m]\setminus\{\delta_{\alpha},\ldots,\delta_{\gamma+1},\delta'_\beta\},\label{samef4}\\
        	&\varepsilon'_h=\varepsilon_h, b'_h=b_h \ \ \text{for any} \ \ h\in[1:\alpha-1]\cup[\beta+1:t],\label{sameb4}\\
        	&\varepsilon'_h=\varepsilon_{h+1},b'_h=b_{h+1}\ \ \text{for any} \ \ h\in[\gamma+1:\beta-1],\label{difb4}
           \end{alignat}
        then for any $h\in[1:\alpha-1]\cup[\beta+1:t]$ we have
            \begin{equation}
            \label{samefeb1}
            \mathbf{P}(f'_{\delta'_h}+(\varepsilon'_h-1)\frac{F_1}{\lambda},b'_h)\overset{\eqref{newT1},\eqref{samef4},\eqref{sameb4}}{=}\mathbf{P}(f_{\delta_h}+(\varepsilon_h-1)\frac{F_1}{\lambda},b_h)=\mathcal{B}_{l_h}[\mu_h]; \end{equation}
        for any $h\in[\gamma+1:\beta-1]$ we have
            \begin{equation}
            \label{difeb}
            \mathbf{P}(f'_{\delta'_h}+(\varepsilon'_h-1)\frac{F_1}{\lambda},b'_h)\overset{\eqref{newT1},\eqref{samef4},\eqref{difb4}}{=}\mathbf{P}(f_{\delta_{h+1}}+(\varepsilon_{h+1}-1)\frac{F_1}{\lambda},b_{h+1})=\mathcal{B}_{l_{h+1}}[\mu_{h+1}].
            \end{equation}
        Assume that $\mathcal{B}_{f_{\delta'_\beta}}[\mu_{\gamma+1}]=s$, there exist $j\in[1:F_1]$ and $k\in[1:K_1]$ such that $\mathbf{P}(j,k)=s$, let $f'_{\delta'_\beta}=\langle j\rangle_{\frac{F_1}{\lambda}}$, $\varepsilon'_\beta=\frac{j-f'_{\delta'_\beta}}{\frac{F_1}{\lambda}}+1$ and $b'_{\beta}=k$, then
            \begin{equation}
            \label{fdeltabeta4}
            \mathbf{P}(f'_{\delta'_{\beta}}+(\varepsilon'_\beta-1)\frac{F_1}{\lambda},b'_{\beta})=\mathcal{B}_{f_{\delta'_\beta}}[\mu_{\gamma+1}].
            \end{equation}
        For any $h\in[\alpha:\gamma]$, assume that $\mathcal{B}_{l_{h+1}}[\mu_{h}]=s_h$, there exist $j_h\in[1:F_1]$ and $k_h\in[1:K_1]$ such that $\mathbf{P}(j_h,k_h)=s_h$, let $f'_{\delta'_{h}}=\langle j_h\rangle_{\frac{F_1}{\lambda}}$, $\varepsilon'_h=\frac{j_h-f'_{\delta'_{h}}}{\frac{F_1}{\lambda}}+1$ and $b'_{h}=k_h$, then
            \begin{equation}
            \label{fbdeltaip2}
            \mathbf{P}(f'_{\delta'_{h}}+(\varepsilon'_h-1)\frac{F_1}{\lambda},b'_{h})=\mathcal{B}_{l_{h+1}}[\mu_{h}],\ \ \text{for any} \ \ h\in[\alpha:\gamma].
            \end{equation}
        From \eqref{samefeb1}, \eqref{difeb}, \eqref{fdeltabeta4} and \eqref{fbdeltaip2}, we have $\mathbf{P}(f'_{\delta'_h}+(\varepsilon'_h-1)\frac{F_1}{\lambda},b'_h)\neq *$ for any $h\in[1:t]$, which implies $\mathbf{P}(f'_{\delta'_h},b'_h)\neq *$ from Condition C4 of Definition \ref{def-basePDA}.
        Let ${\bf f}'=(f'_1,\ldots,f'_m)$, ${\bm \varepsilon}'=(\varepsilon'_1,\ldots,\varepsilon'_t)$, ${\bf b}'=(b'_1,\ldots,b'_t)$, then we have $\mathbf{P}_{m,t}(({\bf f}',{\bm \varepsilon}'),$ $(\mathcal{T}',{\bf b}'))\neq *$ from \eqref{constrPm}. Assume that $\mathbf{P}(f'_{\delta'_h}+(\varepsilon'_h-1)\frac{F_1}{\lambda},b'_h)=\mathcal{B}_{l'_h}[\mu'_h]$ for any $h\in[1:t]$, then from \eqref{samefeb1}, \eqref{difeb}, \eqref{fdeltabeta4} and \eqref{fbdeltaip2} we have
           \begin{equation}
           \label{mu5}
           \mu'_h=\begin{cases}
           \mu_h, \ \ &\text{if} \ \ h\in[1:\gamma]\cup[\beta+1:t],\\
           \mu_{h+1}, \ \ &\text{if} \ \ h\in[\gamma+1:\beta-1],\\
           \mu_{\gamma+1},\ \ &\text{if} \ \ h=\beta
           \end{cases}
           \end{equation}
           and
           \begin{equation}
           \label{l5}
           l'_h=\begin{cases}
           l_h,\ \ &\text{if} \ \ h\in[1:\alpha-1]\cup[\beta+1:t],\\
           l_{h+1}, \ \ &\text{if} \ \ h\in[\alpha:\beta-1],\\
           f_{\delta'_\beta},\ \ &\text{if} \ \ h=\beta.
           \end{cases}
           \end{equation}
         Assume that $\mathbf{P}_{m,t}(({\bf f}',{\bm \varepsilon}'),(\mathcal{T}',{\bf b}'))={\bf e}'=(e'_1,\ldots,e'_m)$, we have
           \begin{alignat*}{10}
           &e'_{\delta_h}\overset{\eqref{newT1}}{=}e'_{\delta'_h}\overset{\eqref{constre}}{=}\mathcal{B}_{l'_h}[\mu'_1]
           \overset{\eqref{mu5},\eqref{l5}}{=}\mathcal{B}_{l_h}[\mu_1]\overset{\eqref{constre}}{=}e_{\delta_h}, h\in[1:\alpha-1], \\
           &e'_{\delta_{h}}\overset{\eqref{newT1}}{=}e'_{\delta'_{h-1}}\overset{\eqref{constre}}{=}\mathcal{B}_{l'_{h-1}}[\mu'_{1}]
           \overset{\eqref{mu5},\eqref{l5}}{=}\mathcal{B}_{l_h}[\mu_1]\overset{\eqref{constre}}{=}e_{\delta_{h}}, h\in[\alpha+1:w],\\
           &e'_{\delta_{h}}\overset{\eqref{newT1}}{=}e'_{\delta'_{h-1}}\overset{\eqref{constre}}{=}\mathcal{B}_{l'_{h-1}}[\mu'_{h-1}]
           \overset{\eqref{mu5},\eqref{l5}}{=}\mathcal{B}_{l_h}[\mu_h]\overset{\eqref{constre}}{=}e_{\delta_{h}}, h\in [w+1:\beta], \\
           &e'_{\delta_{h}}\overset{\eqref{newT1}}{=}e'_{\delta'_{h}}\overset{\eqref{constre}}{=}\mathcal{B}_{l'_{h}}[\mu'_{h}]
           \overset{\eqref{mu5},\eqref{l5}}{=}\mathcal{B}_{l_h}[\mu_h]\overset{\eqref{constre}}{=}e_{\delta_{h}}, h\in [\beta+1:t], \\
           &e'_{\delta_{\alpha}}\overset{\eqref{constre}}{=}\mathcal{B}_{f'_{\delta_{\alpha}}}[\mu'_{1}]
           \overset{\eqref{falfa},\eqref{mu5}}{=}\mathcal{B}_{l_{\alpha}}[\mu_{1}]\overset{\eqref{constre}}{=}e_{\delta_{\alpha}}, \\
           &e'_{\delta'_{\beta}}\overset{\eqref{constre}}{=}\mathcal{B}_{l'_{\beta}}[\mu'_{\beta}]
           \overset{\eqref{mu5},\eqref{l5}}{=}\mathcal{B}_{f_{\delta'_\beta}}[\mu_{\gamma+1}]
           \overset{\eqref{sigmad1}}{=}\mathcal{B}_{f_{\sigma_\gamma}}[\mu_{\gamma+1}]\overset{\eqref{constre}}{=}e_{\sigma_{\gamma}}\overset{\eqref{sigmad1}}{=}e_{\delta'_{\beta}},\\
           &e'_i\overset{\eqref{constre}}{=}\mathcal{B}_{f'_i}[\mu'_1]\overset{\eqref{samef4},\eqref{mu5}}{=}\mathcal{B}_{f_i}[\mu_1]
           \overset{\eqref{constre}}{=}e_i,i\in[1:m-t+1]\setminus (\mathcal{T}\cup\mathcal{T}'),\\
           &e'_{\sigma_h}\overset{\eqref{sigma3}}{=}e'_{\sigma'_h}\overset{\eqref{constre}}{=}\mathcal{B}_{f'_{\sigma'_h}}[\mu'_{h+1}]\overset{\eqref{sigma3},\eqref{samef4},\eqref{mu5}}{=}\mathcal{B}_{f_{\sigma_h}}[\mu_{h+1}]
           \overset{\eqref{constre}}{=}e_{\sigma_h},h\in[1:\gamma-1],\\
           &e'_{\sigma_h}\overset{\eqref{sigma3}}{=}e'_{\sigma'_{h-1}}\overset{\eqref{constre}}{=}\mathcal{B}_{f'_{\sigma'_{h-1}}}[\mu'_{h}]
           \overset{\eqref{sigma3},\eqref{samef4},\eqref{mu5}}{=}\mathcal{B}_{f_{\sigma_h}}[\mu_{h+1}]\overset{\eqref{constre}}{=}e_{\sigma_h},h\in[\gamma+1:w-1],
           \end{alignat*}
           i.e., ${\bf e}'={\bf e}$. Therefore, $\mathbf{P}_{m,t}(({\bf f}',{\bm \varepsilon}'),(\mathcal{T}',{\bf b}'))={\bf e}$.
        If $\alpha>\gamma+1$, we can similarly prove that there exist ${\bf f}'\in[1:\frac{F_1}{\lambda}]^m$, ${\bm \varepsilon}'\in[1:\lambda]^t$ and ${\bf b}'\in[1:K_1]^t$ such that $\mathbf{P}_{m,t}(({\bf f}',{\bm \varepsilon}'),(\mathcal{T}',{\bf b}'))={\bf e}$.

    {\it Case 4:} If $\delta_{\alpha}\in[m-t+2:m]$ and $\delta'_\beta\in[1:m-t+1]$, then $|[m-t+2:m]\setminus\mathcal{T}'|=w$, let $[m-t+2:m]\setminus \mathcal{T}'=\{\sigma'_1,\ldots,\sigma'_{w}\}$ with $\sigma'_1<\ldots<\sigma'_{w}$. Since $\delta_{\alpha}\in\mathcal{T}\cap[m-t+2:m]$ and $[m-t+2:m]\setminus\mathcal{T}=\{\sigma_1,\ldots,\sigma_{w-1}\}$, assume that
        $\sigma_{\gamma}<\delta_{\alpha}<\sigma_{\gamma+1}$, then we have \eqref{newT} and
        \begin{equation}
        \label{sigma2}
        \sigma'_h=\begin{cases}
        \sigma_h, \ \ &\text{if} \ \ h\in[1:\gamma],\\
        \delta_{\alpha}, \ \ &\text{if} \ \  h=\gamma+1, \\
        \sigma_{h-1}, \ \ &\text{if} \ \  h\in[\gamma+2:w].
        \end{cases}
        \end{equation}
        If $\beta>\gamma+2$, let
            \begin{alignat}{3}
            &f'_i=f_i \ \ \text{for any} \ \ i\in[1:m]\setminus\{\delta_{\gamma+2},\ldots,\delta_{\beta-1},\delta'_\beta,\delta_{\alpha}\},\label{samef5}\\
            &\varepsilon'_h=\varepsilon_h, b'_h=b_h \ \ \text{for any} \ \ h\in[1:\gamma+1]\cup[\alpha+1:t],\label{sameb5}\\
            &\varepsilon'_h=\varepsilon_{h-1}, b'_h=b_{h-1}\ \ \text{for any} \ \ i\in[\beta+1:\alpha],\label{difb5}
            \end{alignat}
            then for any $h\in[1:\gamma+1]\cup[\alpha+1:t]$ we have
            \begin{equation}
            \label{samefeb2}
            \mathbf{P}(f'_{\delta'_h}+(\varepsilon'_h-1)\frac{F_1}{\lambda},b'_h)\overset{\eqref{newT},\eqref{samef5},\eqref{sameb5}}{=}\mathbf{P}(f_{\delta_h}+(\varepsilon_h-1)\frac{F_1}{\lambda},b_h)=\mathcal{B}_{l_h}[\mu_h];
            \end{equation}
            for any $h\in[\beta+1:\alpha]$ we have
            \begin{equation}
            \label{safdib}
            \mathbf{P}(f'_{\delta'_h}+(\varepsilon'_h-1)\frac{F_1}{\lambda},b'_h)\overset{\eqref{newT},\eqref{samef5},\eqref{difb5}}{=}\mathbf{P}(f_{\delta_{h-1}}+(\varepsilon_{h-1}-1)\frac{F_1}{\lambda},b_{h-1})=\mathcal{B}_{l_{h-1}}[\mu_{h-1}].
            \end{equation}
            Assume that $\mathcal{B}_{l_{\gamma+2}}[\mu_{\alpha}]=s$, there exist $j\in[1:F_1]$ and $k\in[1:K_1]$ such that $\mathbf{P}(j,k)=s$, let $f'_{\delta'_{\gamma+2}}=\langle j\rangle_{\frac{F_1}{\lambda}}$, $\varepsilon'_{\gamma+2}=\frac{j-f'_{\delta'_{\gamma+2}}}{\frac{F_1}{\lambda}}+1$ and $b'_{\gamma+2}=k$, then
            \begin{equation}
            \label{fbgamap1}
            \mathbf{P}(f'_{\delta'_{\gamma+2}}+(\varepsilon'_{\gamma+2}-1)\frac{F_1}{\lambda},b'_{\gamma+2})=\mathcal{B}_{l_{\gamma+2}}[\mu_{\alpha}].
            \end{equation}
            For any $h\in[\gamma+3:\beta-1]$, assume that $\mathcal{B}_{l_h}[\mu_{h-1}]=s_h$, there exist $j_h\in[1:F_1]$ and $k_h\in[1:K_1]$ such that $\mathbf{P}(j_h,k_h)=s_h$, let $f'_{\delta'_{h}}=\langle j_h\rangle_{\frac{F_1}{\lambda}}$, $\varepsilon'_h=\frac{j_h-f'_{\delta'_{h}}}{\frac{F_1}{\lambda}}+1$ and $b'_{h}=k_h$, then
            \begin{equation}
            \label{fbdeltaip1}
            \mathbf{P}(f'_{\delta'_{h}}+(\varepsilon'_h-1)\frac{F_1}{\lambda},b'_{h})=\mathcal{B}_{l_h}[\mu_{h-1}], \ h\in[\gamma+3:\beta-1].
            \end{equation}
            Assume that $\mathcal{B}_{f_{\delta'_{\beta}}}[\mu_{\beta-1}]=s'$, there exist $j'\in[1:F_1]$ and $k'\in[1:K_1]$ such that $\mathbf{P}(j',k')=s'$, let $f'_{\delta'_\beta}=\langle j'\rangle_{\frac{F_1}{\lambda}}$, $\varepsilon'_\beta=\frac{j'-f'_{\delta'_\beta}}{\frac{F_1}{\lambda}}+1$ and $b'_{\beta}=k'$, then
                \begin{equation}
                \label{fbbetap}
                \mathbf{P}(f'_{\delta'_{\beta}}+(\varepsilon'_\beta-1)\frac{F_1}{\lambda},b'_{\beta})=\mathcal{B}_{f_{\delta'_{\beta}}}[\mu_{\beta-1}].
                \end{equation}
            From \eqref{samefeb2}, \eqref{safdib}, \eqref{fbgamap1}, \eqref{fbdeltaip1} and \eqref{fbbetap} we have $\mathbf{P}(f'_{\delta'_{h}}+(\varepsilon'_h-1)\frac{F_1}{\lambda},b'_{h})\neq *$ for any $h\in[1:t]$, which implies that $\mathbf{P}(f'_{\delta'_h},b'_h)\neq*$ from Condition C4 of Definition \ref{def-basePDA}.
            Let ${\bf f}'=(f'_1,\ldots,f'_m)$, ${\bm \varepsilon}'=(\varepsilon'_1,\ldots,\varepsilon'_t)$, ${\bf b}'=(b'_1,\ldots,b'_t)$, then we have $\mathbf{P}_{m,t}(({\bf f}',{\bm \varepsilon}'),(\mathcal{T}',{\bf b}'))\neq *$ from \eqref{constrPm}. Assume that $\mathbf{P}(f'_{\delta'_h}+(\varepsilon'_h-1)\frac{F_1}{\lambda},b'_h)=\mathcal{B}_{l'_h}[\mu'_h]$ for any $h\in[1:t]$, then from \eqref{samefeb2}, \eqref{safdib}, \eqref{fbgamap1}, \eqref{fbdeltaip1} and \eqref{fbbetap} we have
           \begin{equation}
           \label{mu6}
           \mu'_h=\begin{cases}
           \mu_h, \ \ &\text{if} \ \ h\in[1:\gamma+1]\cup[\alpha+1:t],\\
           \mu_{\alpha},\ \ &\text{if} \ \ h=\gamma+2,\\
           \mu_{h-1}, \ \ &\text{if} \ \ h\in[\gamma+3:\alpha].
           \end{cases}
           \end{equation}
           and
           \begin{equation}
           \label{l6}
           l'_h=\begin{cases}
           l_h,\ \ &\text{if} \ \ h\in[1:\beta-1]\cup[\alpha+1:t],\\
           f_{\delta'_\beta},\ \ &\text{if} \ \ h=\beta,\\
           l_{h-1}, \ \ &\text{if} \ \ h\in[\beta+1:\alpha].
           \end{cases}
           \end{equation}
         Assume that $\mathbf{P}_{m,t}(({\bf f}',{\bm \varepsilon}'),(\mathcal{T}',{\bf b}'))={\bf e}'=(e'_1,\ldots,e'_m)$, we have
           \begin{alignat*}{10}
           &e'_{\delta_h}\overset{\eqref{newT}}{=}e'_{\delta'_h}\overset{\eqref{constre}}{=}\mathcal{B}_{l'_h}[\mu'_1]
           \overset{\eqref{mu6},\eqref{l6}}{=}\mathcal{B}_{l_h}[\mu_1]\overset{\eqref{constre}}{=}e_{\delta_h},h\in[1:\beta-1],\\
           &e'_{\delta_{h}}\overset{\eqref{newT}}{=}e'_{\delta'_{h+1}}\overset{\eqref{constre}}{=}\mathcal{B}_{l'_{h+1}}[\mu'_{1}]
           \overset{\eqref{mu6},\eqref{l6}}{=}\mathcal{B}_{l_h}[\mu_1]\overset{\eqref{constre}}{=}e_{\delta_{h}}, h\in[\beta:w], \\
           &e'_{\delta_{h}}\overset{\eqref{newT}}{=}e'_{\delta'_{h+1}}\overset{\eqref{constre}}{=}\mathcal{B}_{l'_{h+1}}[\mu'_{h+1}]
           \overset{\eqref{mu6},\eqref{l6}}{=}\mathcal{B}_{l_h}[\mu_h]\overset{\eqref{constre}}{=}e_{\delta_{h}}, h\in [w+1:\alpha-1],\\
           &e'_{\delta_{h}}\overset{\eqref{newT}}{=}e'_{\delta'_{h}}\overset{\eqref{constre}}{=}\mathcal{B}_{l'_{h}}[\mu'_{h}]
           \overset{\eqref{mu6},\eqref{l6}}{=}\mathcal{B}_{l_h}[\mu_h]\overset{\eqref{constre}}{=}e_{\delta_{h}}, h\in [\alpha+1:t],\\
           &e'_{\delta_{\alpha}}\overset{\eqref{sigma2}}{=}e'_{\sigma'_{\gamma+1}}\overset{\eqref{constre}}{=}\mathcal{B}_{f'_{\sigma'_{\gamma+1}}}[\mu'_{\gamma+2}]\overset{\eqref{sigma2}}{=}\mathcal{B}_{f'_{\delta_\alpha}}[\mu'_{\gamma+2}]
           \overset{\eqref{falfa},\eqref{mu6}}{=}\mathcal{B}_{l_{\alpha}}[\mu_{\alpha}]\overset{\eqref{constre}}{=}e_{\delta_{\alpha}},           
           \end{alignat*}
       \begin{alignat*}{10}
       	&e'_{\delta'_\beta}\overset{\eqref{constre}}{=}\mathcal{B}_{l'_{\beta}}[\mu'_{1}]\overset{\eqref{mu6},\eqref{l6}}{=}\mathcal{B}_{f_{\delta'_\beta}}[\mu_{1}]
       	\overset{\eqref{constre}}{=}e_{\delta'_\beta},\\
       	&e'_i\overset{\eqref{constre}}{=}\mathcal{B}_{f'_i}[\mu'_1]\overset{\eqref{samef5},\eqref{mu6}}{=}\mathcal{B}_{f_i}[\mu_1]
       	\overset{\eqref{constre}}{=}e_i, i\in[1:m-t+1]\setminus (\mathcal{T}\cup\mathcal{T}'),\\
       	&e'_{\sigma_h}\overset{\eqref{sigma2}}{=}e'_{\sigma'_h}\overset{\eqref{constre}}{=}\mathcal{B}_{f'_{\sigma'_h}}[\mu'_{h+1}]\overset{\eqref{sigma2}}{=}\mathcal{B}_{f'_{\sigma_h}}[\mu'_{h+1}]
       	\overset{\eqref{samef5},\eqref{mu6}}{=}\mathcal{B}_{f_{\sigma_h}}[\mu_{h+1}]\overset{\eqref{constre}}{=}e_{\sigma_h},h\in[1:\gamma],\\
       	&e'_{\sigma_h}\overset{\eqref{sigma2}}{=}e'_{\sigma'_{h+1}}\overset{\eqref{constre}}{=}\mathcal{B}_{f'_{\sigma'_{h+1}}}[\mu'_{h+2}]
       	\overset{\eqref{sigma2},\eqref{samef5},\eqref{mu6}}{=}\mathcal{B}_{f_{\sigma_h}}[\mu_{h+1}]        \overset{\eqref{constre}}{=}e_{\sigma_h},h\in[\gamma+1:w-1],
       \end{alignat*}
           i.e., ${\bf e}'={\bf e}$. Therefore, $\mathbf{P}_{m,t}(({\bf f}',{\bm \varepsilon}'),(\mathcal{T}',{\bf b}'))={\bf e}$.
       If $\beta<\gamma+2$ or $\beta=\gamma+2$, we can similarly prove that there exist ${\bf f}'\in[1:\frac{F_1}{\lambda}]^m$, ${\bm \varepsilon}'\in[1:\lambda]^t$ and ${\bf b}'\in[1:K_1]^t$ such that $\mathbf{P}_{m,t}(({\bf f}',{\bm \varepsilon}'),(\mathcal{T}',{\bf b}'))={\bf e}$.


    If $|\mathcal{T}'\cap \mathcal{T}|<t-1$, we can find a sequence $\mathcal{T}_1,\mathcal{T}_2,\ldots,\mathcal{T}_k\in{[1:m]\choose t}$ such that $\mathcal{T}_1=\mathcal{T}$, $\mathcal{T}_k=\mathcal{T}'$ and $|\mathcal{T}_{i}\cap\mathcal{T}_{i+1}|=t-1$ for $i\in[1:k-1]$. So there exist ${\bf f}'\in[1:\frac{F_1}{\lambda}]^m$, ${\bm \varepsilon}'\in[1:\lambda]^t$ and ${\bf b}'\in[1:K_1]^t$ such that $\mathbf{P}_{m,t}(({\bf f}',{\bm \varepsilon}'),(\mathcal{T}',{\bf b}'))={\bf e}$. The first statement holds.

    Now let us consider the second statement.
    For any $h\in[1:t]$, we have $\mathbf{P}(f_{\delta_h},b_h)\neq *$ from \eqref{constrPm}, which implies $\mathbf{P}(f_{\delta_h}+(\varepsilon_h-1)\frac{F_1}{\lambda},b_h)\neq *$ from Condition C4 of Definition \ref{def-basePDA}. Assume that $\mathbf{P}(f_{\delta_h}+(\varepsilon_h-1)\frac{F_1}{\lambda},b_h)=s_h$, then $s_h$ appears exactly $g_1$ times in $\mathbf{P}$, since $\mathbf{P}$ is a $g_1$-PDA. Assume that $\mathbf{P}(j_{s_h,v},k_{s_h,v})=s_h$ where $v\in[1:g_1]$.
    Let $f'_i=f_i$ for any $i\in[1:m]\setminus\mathcal{T}$. For any $h\in[1:t]$, let $f'_{\delta_h}=\langle j_{s_h,v}\rangle_{\frac{F_1}{\lambda}}$, $\varepsilon'_{h}=\frac{j_{s_h,v}-f'_{\delta_h}}{\frac{F_1}{\lambda}}+1$ and $b'_h=k_{s_h,v}$ where $v\in[1:g_1]$, then we have $\mathbf{P}(f'_{\delta_h}+(\varepsilon'_h-1)\frac{F_1}{\lambda},b'_h)=s_h=\mathcal{B}_{l_h}[\mu_h]$. Let ${\bf f}'=(f'_1,\ldots,f'_m)$, ${\bm \varepsilon'}=(\varepsilon'_1,\ldots,\varepsilon'_t)$ and ${\bf b}'=(b'_1,\ldots,b'_t)$, then we have $\mathbf{P}_{m,t}(({\bf f}',{\bm \varepsilon}'),(\mathcal{T},{\bf b}'))=\mathbf{P}_{m,t}(({\bf f},{\bm \varepsilon}),(\mathcal{T},{\bf b}))={\bf e}$ from \eqref{constre}. Hence, ${\bf e}$ appears exactly $g^t$ times in the columns indexed by $(\mathcal{T},{\bf c})$ where ${\bf c}\in [1:K_1]^t$. The second statement holds.

    The two statements imply that each vector ${\bf e}$ in $\mathbf{P}_{m,t}$ appears exactly ${m\choose t}g_1^t$ times, hence, we have $S=\frac{{m\choose t}K_1^t\lambda^t(\frac{F_1}{\lambda})^{m}(F_1-Z_1)^t/F_1^t}{{m\choose t}g_1^t}=(\frac{F_1}{\lambda})^{m-t}S_1^t$, since $g_1=\frac{K_1(F_1-Z_1)}{S_1}$. The condition C$2$ of Definition \ref{def-PDA} holds.

Therefore, the array $\mathbf{P}_{m,t}$ generated by Construction \ref{constr1} is an ${m\choose t}g^t$-$\big({m\choose t}K_1^t,\lambda^t (\frac{F_1}{\lambda})^m,\lambda^t (\frac{F_1}{\lambda})^m(1-\big.$  $\big.(\frac{F_1-Z_1}{F_1})^t),(\frac{F_1}{\lambda})^{m-t}S_1^t\big)$ PDA, which generates a $(K,M,N)$ coded caching scheme with the number of users $K={m\choose t}K_1^t$, memory ratio $\frac{M}{N}=1-\big(\frac{F_1-Z_1}{F_1}\big)^t$, subpacketization $F=\lambda^t (\frac{F_1}{\lambda})^m$, and load $R=\big(\frac{S_1}{F_1}\big)^t$ from Lemma \ref{th-Fundamental}.
\end{proof}

\bibliographystyle{IEEEtran}
\bibliography{reference}

\end{document}